\newcommand{\printstyle}{reprint}
\newcommand{\halfwidth}{\columnwidth} 

\newcommand{\thirdwidth}{0.9\halfwidth}

\documentclass[aip,pop,floatfix,\printstyle ,nofootinbib]{revtex4-1}

\usepackage{color}

\usepackage[caption=false]{subfig}
\usepackage{graphicx,booktabs,array}
\usepackage{multirow}
\DeclareGraphicsExtensions{.pdf,.png,.jpg,.eps}
\usepackage{epsfig}
\usepackage{epstopdf}
\usepackage{amssymb}
\usepackage{amsmath}
\usepackage{amsfonts}
\usepackage{mathrsfs}
\usepackage{amsthm}
\usepackage{float}
\usepackage{amsbsy}
\usepackage{bm}
\usepackage{grffile}
\usepackage{hyperref}
\hypersetup{
    colorlinks = true,
    citecolor = blue,
    urlcolor = blue,
    linkcolor = blue,
}
\usepackage{courier}
\usepackage{upgreek}
\usepackage{xspace}
\usepackage{xcolor}
\usepackage{calc} 
\usepackage{soul} 

\newcommand{\ie}{\textit{i}.\textit{e}. }
\newcommand{\eg}{\textit{e}.\textit{g}. }
\newcommand{\etc}{\textit{etc.}\xspace} 

\newcommand{\Alfven}{Alfv\'{e}n\xspace}
\newcommand{\Alfvenic}{Alfv\'{e}nic\xspace}

\newcommand{\va}{v_A}

\newcommand{\vc}{v_c}

\newcommand{\omegab}{\omega_b} 

\newcommand{\neo}{n_e}

\newcommand{\vpar}{v_\parallel}
\newcommand{\vpres}{v_{\parallel,\text{res}}}

\newcommand{\vtherme}{v_{th,e}}
\newcommand{\vthermi}{v_{th,i}}
\newcommand{\tep}{\text{EP}}
\newcommand{\vep}{v_\tep}
\newcommand{\Eep}{\W_\tep}

\newcommand{\W}{\mathcal{E}}

\newcommand{\Te}{T_e}

\newcommand{\nuhat}{\hat{\nu}}
\newcommand{\alphahat}{\hat{\alpha}}

\newcommand{\Eamp}{\hat{E}}
\newcommand{\Ehat}{\Eamp(t)}

\newcommand{\gammal}{\gamma_{L}}
\newcommand{\gammanl}{\gamma_{NL}}
\newcommand{\gammad}{\gamma_{d}}
\newcommand{\br}{\Re{b}}
\newcommand{\bi}{\Im{b}}
\newcommand{\brvar}{\Re{b(\alphahat,\nuhat)}}
\newcommand{\Aamp}{\abs{A(\tau)}}
\newcommand{\Aampz}{\abs{A_0}}
\newcommand{\Aexact}{A_\text{exact}}
\newcommand{\phit}{\phi(\tau)}
\newcommand{\phiz}{\phi_0}

\newcommand{\eulergamma}{\gamma_E}
\newcommand{\Asat}{A_\text{sat}}
\newcommand{\omegabsat}{\omega_{b,\text{sat}}}
\newcommand{\omegabsq}{\abs{\omegab^2}}
\newcommand{\desat}{\delta E_\text{sat}}

\newcommand{\tinfl}{\tau_\text{infl}}

\newcommand{\tdiv}{\tau_\text{div}}
\newcommand{\vres}{v_\text{res}}

\newcommand{\rrat}{\alpha/\nu}

\newcommand{\domegasat}{\delta\omega_\text{sat}}

\newcommand{\ff}{F}
\newcommand{\fz}{F_0}
\newcommand{\df}{\delta F}

\newcommand{\taus}{\tau_s}
\newcommand{\tausinv}{\taus^{-1}}
\newcommand{\nuperp}{\nu_\perp}

\newcommand{\vresl}{v_{\text{res},L}}
\newcommand{\vresnl}{v_{\text{res},NL}}
\newcommand{\dvres}{\delta\vres}
\newcommand{\omegal}{\omega_L}

\newcommand{\pitch}{\chi}
\newcommand{\Wres}{\W_\text{res}}
\newcommand{\mpro}{m_p}
\newcommand{\gdgl}{\gammad/\gammal}

\newcommand{\avg}[1]{\left\langle #1 \right\rangle}

\newcommand{\ten}[1]{\cdot 10^{#1}}
\newcommand{\abs}[1]{\left|#1\right|}

\newcommand{\defined}{\equiv}
\newcommand{\like}{\sim}

\newcommand{\ord}[1]{\mathcal{O}\left(#1\right)}
\newcommand{\plusord}[1]{\, + \, \ord{#1}}

\newcommand{\approptoinn}[2]{\mathrel{\vcenter{
  \offinterlineskip\halign{\hfil$##$\cr
    #1\propto\cr\noalign{\kern2pt}#1\sim\cr\noalign{\kern-2pt}}}}}

\newcommand{\tto}{\text{ to }}

\newcommand\numberthis{\addtocounter{equation}{1}\tag{\theequation}}
\renewcommand{\Re}[1]{\text{Re}\left[#1\right]}
\renewcommand{\Im}[1]{\text{Im}\left[#1\right]}

\newcommand{\nofrac}[2]{\genfrac{}{}{0pt}{}{#1}{#2}}
\newcommand{\MG}[9]{G_{#3,#4}^{#1,#2}\left(#9 \left| \nofrac{#5,#6}{#7,#8}\right) \right.}
\newcommand{\MGsimp}[5]{G_{0,0}^{#1,#2}\left(#5 \left| \nofrac{#3}{#4}\right) \right.}

\newcommand{\pderiv}[2]{\frac{\partial #1}{\partial #2}}

\newcommand{\deriv}[2]{\frac{d #1}{d #2}}

\newcommand{\figref}[1]{Fig.\xspace\ref{#1}}
\renewcommand{\eqref}[1]{Eq.\xspace\ref{#1}}
\newcommand{\secref}[1]{Sec.\xspace\ref{#1}}
\newcommand{\citeref}[1]{Ref.\xspace\onlinecite{#1}}
\newcommand{\appref}[1]{Appendix\xspace\ref{#1}}

\newcommand{\code}[1]{\texttt{#1}\xspace}

\newcommand{\NOVAK}{\code{NOVA-K}}

\newcommand{\BOT}{\code{BOT}}
\newcommand{\Mathematica}{\code{Mathematica}}

\newcommand{\myname}{J.B. Lestz}

\newcommand{\Vinicius}{V.N. Duarte}

\newcommand{\PPPLlong}{Princeton Plasma Physics Laboratory, Princeton University, Princeton, New Jersey 08543, USA}

\newcommand{\UCIrvine}{Department of Physics and Astronomy, University of California, Irvine, CA 92697, USA}
\newcommand{\UCI}{\UCIrvine}

\newcommand{\Podesta}{Podest{\`a}}

\definecolor{darkgreen}{rgb}{0,0.5,0}

\begin{document}

\title{Analytic quasi-steady evolution of a marginally unstable wave in the presence of drag and scattering}
\author{\myname}
\email{jlestz@uci.edu}
\affiliation{\UCI}
\author{\Vinicius}
\affiliation{\PPPLlong}
\date{\today}
\begin{abstract}
The 1D bump-on-tail problem is studied in order to determine the influence of drag on quasi-steady solutions near marginal stability ($1-\gdgl \ll 1$) when effective collisions are much larger than the instability growth rate ($\nu \gg \gamma$). In this common tokamak regime, it is rigorously shown that the paradigmatic Berk-Breizman cubic equation for the nonlinear mode evolution reduces to a much simpler differential equation, dubbed the time-local cubic equation, which can be solved directly. It is found that in addition to increasing the saturation amplitude, drag introduces a shift in the apparent oscillation frequency by modulating the saturated wave envelope. Excellent agreement is found between the analytic solution for the mode evolution and both the numerically integrated Berk-Breizman cubic equation and fully nonlinear 1D Vlasov simulations. Experimentally isolating the contribution of drag to the saturated mode amplitude for verification purposes is explored but complicated by the reality that the amount of drag can not be varied independently of other key parameters in realistic scenarios. While the effect of drag is modest when the ratio of drag to scattering $\rrat$ is very small, it can become substantial when $\rrat \gtrsim 0.5$, suggesting that drag should be accounted for in quantitative models of fast-ion-driven instabilities in fusion plasmas. 
\end{abstract}
\maketitle

\section{Introduction}
\label{sec:introduction}

Wave-particle interactions are responsible for a diverse array of phenomena in plasmas. In fusion experiments, energetic particles can drive MHD modes unstable, which in turn can transport those energetic particles, jeopardizing fusion performance and presenting a hazard to the plasma-facing components.\cite{Fasoli2007NF,Heidbrink2008POP,Sharapov2013NF,Gorelenkov2014NF,Heidbrink2020POP} While the linear theory of these instabilities is now well developed, fundamental features of the nonlinear dynamics of such systems are still not known. Understanding their nonlinear evolution is required to interpret experimental observations, which almost exclusively occur in the nonlinear regime, as well as build intuition to guide the development of predictive models of fast ion transport in fusion devices, which is an outstanding need for upcoming burning plasma experiments. A hallmark of nonlinear systems is that even simple systems can contain rich dynamics. Hence, studying basic models can provide insight into the more complex systems which exist in laboratory or naturally occurring plasmas. 

In this work, we seek to understand how drag (also known as dynamical friction) influences the nonlinear evolution of an isolated eigenmode driven unstable by fast ions in the quasi-steady regime. The non-resonant effect of drag on an energetic beam injected into a plasma leads to the formation of a slowing down distribution,\cite{Gaffey1976JPP} while diffusive scattering results in pitch angle broadening.\cite{Cordey1974PF} It is now recognized that the competition between convective drag and diffusive scattering collisions within wave-particle resonances plays a key role in determining whether an instability will exhibit quasi-steady behavior (\ie oscillate at a fixed frequency with a roughly constant amplitude) or one of a variety of non-steady/dynamical phenomena (such as frequency chirping, bursting, chaos, \etc).\cite{Lilley2010POP,Lesur2010POP,Lesur2012NF,Duarte2017NF,Duarte2017POP,Duarte2018NF,Hou2018NF,Slaby2019NF} However, the effects of drag on the more ubiquitous, quasi-steady solutions has not been studied as extensively, despite the fact that drag is a component of the widely used Fokker-Planck collision operator for fast ions.\cite{Rosenbluth1957PRL,Trubnikov1965RPP,Goldston1981JCP} 

The bump-on-tail problem is a common starting point for deriving nonlinear saturation levels. \cite{Fried1971Report,Dewar1973PF,Berk1990PFB1,Berk1990PFB2,Berk1990PFB3} It describes a resonant interaction between a plasma wave and an energetic minority species with free energy available to drive the wave unstable due to a gradient in its inverted distribution function. Near marginal stability, the mode evolution obeys the Berk-Breizman cubic (BBC) equation, which is a nonlinear, time-delayed integro-differential equation.\cite{Berk1996PRL,Berk1997PPR,Berk1997report,Breizman1997POP,Lilley2009PRL} Overviews of the Berk-Breizman theory of weakly-nonlinear instabilities can be found in \citeref{Breizman2011PPCF} and \citeref{Todo2019RMPP}, with the latter adopting a pedagogical approach. There does not exist an analogous nonlinear equation prescribing the general evolution of the mode amplitude far from marginal stability (strongly-nonlinear regime) which is decoupled from the simultaneous evolution of the particle distribution. It was recently shown that for either diffusive or Krook (creation/annihilation) collisions that are sufficiently large (relative to the growth rate) within a narrow resonance, the BBC equation admits an explicit solution for the evolution of the mode.\cite{Duarte2019NF} The present work builds upon this insight, fortifying its mathematical foundations and also generalizing the solution to include the effect of drag. 

Hence, a new analytic solution to the 1D electrostatic bump-on-tail problem is presented in this paper which takes into account both the effect of drag and scattering. This quasi-steady solution is derived from a new nonlinear evolution equation, dubbed the time-local cubic (TLC) equation, which is applicable when the mode is sufficiently close to marginal stability and the effective rate of collisions within the resonance are sufficiently large. Both of these conditions are routinely satisfied in fusion plasmas. Not only does this solution reproduce the previously known destabilizing effect of drag on the saturation level,\cite{Lilley2009thesis,Lesur2010POP} but importantly it describes the entire evolution of the mode from linear growth through nonlinear saturation. This is similar to the solution that was recently derived for the case of scattering only,\cite{Duarte2019NF} though drag introduces both quantitative corrections and qualitative new features. In particular, finite drag acts to evolve the phase of the complex mode amplitude (whereas only its magnitude is evolved without drag), introducing an apparent frequency shift in the overall fluctuations. The quantitative effect of drag on the mode evolution is limited until drag becomes at least half as large as the rate of effective scattering, at which point the saturation amplitude and frequency shift become very sensitive to the ratio of drag to scattering. It is worth mentioning that the evolution of the mode discussed in this paper is only one half of the story -- the other half is the evolution of the resonant particle distribution. In \citeref{Duarte2021U}, similar methods are used to demonstrate that the distribution function satisfies a quasilinear diffusion-advection equation with shifted resonance lines due to drag. 

Special attention is given in the present work to establishing the validity of the time-local approach in order to determine when it applies as a function of the rates of scattering and drag collisions, wave damping, and wave drive. The resulting constraints indicate when the simplified formulas can be reliably used to gain understanding of nonlinear phenomena, especially with regards to experimental interpretation, versus when it becomes necessary to resort to more expensive and mathematically opaque first principles simulations. Comparison with numerically integrated solutions of the BBC equation confirms that they converge to the analytic TLC solution in the anticipated regime. A transition from irregular behavior in numerical BBC solutions to the smooth, quasi-steady solutions is observed as the input parameters are varied across the threshold for validity of the TLC equation. The analytic TLC solution is also compared to fully nonlinear 1D Vlasov simulations, finding similarly strong agreement. Unfortunately, predictions of the measurable effect of drag on tokamak instabilities are difficult to formulate due to the fact that realistically, the key parameters of the TLC solution can not be controlled independently due to the interconnected way in which they depend on underlying properties of the background plasma. 

This paper is organized as follows. In \secref{sec:analytic}, the time-local cubic (TLC) equation with drag and scattering is derived from the Berk-Breizman cubic (BBC) equation. The features and validity of its exact analytic solution are further discussed in the subsections of \secref{sec:analytic}. In \secref{sec:cubic}, the analytic TLC solution is compared to numerically integrated solutions of the BBC. The dependence of the predicted saturation amplitude and frequency shift on the ratio of drag to scattering is compared against fully nonlinear 1D Vlasov simulations in \secref{sec:bot}. The potential experimental signatures of drag on properties of toroidal \Alfven eigenmodes are considered in \secref{sec:experiment}. Lastly, a summary of the main results and closing discussion is given in \secref{sec:conclusions}. 

\section{Analytic evolution with large effective scattering near marginal stability \texorpdfstring{$(\nu \gg \gamma,\omegab)$}{(nu >> gamma,omega\_b)}}
\label{sec:analytic}

\subsection{Derivation of the time-local cubic (TLC) equation}
\label{sec:exact}

The starting point for this work is the well-studied 1D electrostatic bump-on-tail problem, using a collision operator that includes both diffusive scattering and convective drag. In this section, we will show that the nonlinear evolution of the complex mode amplitude obeys a much simpler ordinary differential equation when the effective scattering rate is much larger than both 1) the net growth rate of the mode $(\nu \gg \gamma)$ and 2) the unperturbed (collisionless) phase space bounce frequency of deeply trapped resonant particles $(\nu^2 \gg \omegabsq)$. These conditions will be more precisely quantified in \secref{sec:validity}, with a detailed derivation given in \appref{app:rval}. 

The kinetic equation for this bump-on-tail problem is the following Vlasov system

\begin{align}
\pderiv{\ff}{t} + v\pderiv{\ff}{x} + \frac{q_i E(x,t)}{m_i}\pderiv{\ff}{v}
= \frac{\nu^3}{k^2}\pderiv{^2\df}{v^2} + \frac{\alpha^2}{k}\pderiv{\df}{v}
\label{eq:vlasov}
\end{align}

Here, $\ff(x,v,t)$ is the distribution function for a minority population of high energy ions with mass $m_i$ and charge $q_i$. The distribution is decomposed into its equilibrium and fluctuating parts: $\ff(x,v,t) = \fz(v) + \df(x,v,t)$. We will assume that $\fz(v)$ has a region with $\partial\fz/\partial v > 0$, capable of driving a mode unstable via inverse Landau damping.\cite{Landau1946,Kolesnichenko1967SAE,Rosenbluth1975PRL,Mikhailovskiiv6} The electric field perturbation is assumed to be a monochromatic wave $E(x,t) = \Re{\Ehat e^{i(kx-\omega t)}}$ where $\Ehat$ is the complex mode amplitude which we want to determine as a function of time. It is assumed that the linear frequency $\omega$ and wavenumber $k$ of the mode are chosen such that the resonant phase velocity $\vres = \omega/k$ lies within the high energy bump in the distribution. Moreover, only a narrow range of velocities near $\vres$ is analyzed, such that the slope of $\fz$ is approximately constant in this region and proportional to the linear fast ion drive at $t = 0$: $\gammal =  2\pi^2(q_i^2\omega/m_i k^2)\left.\fz'(v)\right|_{v=\vres}$. The mode's damping rate due to all interactions with the thermal plasma is given by a constant $\gammad$. 

The collision operator on the right-hand-side of \eqref{eq:vlasov} contains both diffusive scattering with frequency $\nu$ and convective drag (also referred to as dynamical friction) with frequency $\alpha$. The specific form follows from the Fokker-Planck collision operator in the limit of $\vthermi \ll \vep \ll \vtherme$.\cite{Trubnikov1965RPP,Berk1975NF,Berk1997PPR,Berk1997report} It is important to note that the collisional coefficients $\nu$ and $\alpha$ appearing in \eqref{eq:vlasov} are the \emph{effective} rates of scattering and drag collisions within a narrow resonance in phase space. Hence they are enhanced relative to the familiar $90^\circ$ pitch angle scattering rate $\nu_\perp$ and inverse slowing down time $\tausinv$, respectively. The effective collision frequencies within the resonance scale as $\nu \like \nu_\perp \omega^2 / (k\Delta v)^2$ and $\alpha \like \tau_s^{-1}\omega/(k\Delta v)$, where $k\Delta v \like \nu,\alpha$ is the characteristic collisional broadening of the resonance. In other words, the effective collision frequencies scale as $\nu \like \left(\nu_\perp \omega^2\right)^{1/3}$ and $\alpha \like \left(\omega/\tau_s\right)^{1/2}$. This enhancement occurs because the resonant particle dynamics are sensitive to the timescale of collisional scattering across the narrow resonant region, which is much faster than the typical scattering rate across phase space in general.\cite{Su1968PRL,Berk1992PRL,Callen2014POP,Catto2020JPP} Since $\nuperp \like \neo/\Eep^{3/2}$ and $\taus \like \Te^{3/2}/\neo$,  the heuristic scaling of the ratio of drag to scattering is given by 

\begin{align}
\frac{\alpha}{\nu} \like \frac{1}{\taus^{1/2}\nuperp^{1/3}\omega^{1/6}} \like \frac{\Eep^{1/2}\neo^{1/6}}{\Te^{3/4}\omega^{1/6}}
\label{eq:alphaovernu-heuristic}
\end{align}

Here, $n_e$ and $T_e$ are the electron density and temperature while $\Eep$ is the energy of resonant particles. Moreover, microturbulence can further increase $\nu$ and affect the ratio of drag to scattering.\cite{Lang2011POP,Duarte2017NF,Duarte2017POP,Duarte2018NF} More detailed expressions for $\nu$ and $\alpha$ for passing particles in a tokamak have previously been derived,\cite{Lilley2009PRL,Lesur2010POP,Duarte2017POP} but the heuristic expression in \eqref{eq:alphaovernu-heuristic} is sufficient for the qualitative analysis of this paper, and some of its experimental implications will be explored in \secref{sec:experiment}.

With suitable initial and boundary conditions, the partial differential equation in \eqref{eq:vlasov} can be combined with a power balance equation (describing the energy transfer to the fast ions and background plasma) in order to determine the full evolution of the electric field perturbation $\Ehat$ and particle distribution $\df(x,v,t)$. However, these quantities are nonlinearly coupled together and no general solution exists for this system. Fortunately, conditions have been found which decouple these two quantities of interest, allowing for further analytic progress to be made. 

In the absence of collisions, resonantly trapped particles follow elliptic orbits in phase space due to the electrostatic wave potential. This periodic motion occurs with a bounce frequency defined by the magnitude of $\omegab^2(t) = q_i k \Ehat / m_i$. When $\nu^2 \gg \omegabsq$, many collisions occur within a single orbit period, destroying the orbital trajectories and replacing them with highly disorganized particle motion within the resonant island. In this regime, $\df$ can be Fourier expanded in successive powers of $\omegabsq/\nu^2 \ll 1$, enabling analytic progress. The loss of phase correlations within the island due to frequent collisions makes it possible to decouple $\Ehat$ from $\df$ so that the mode evolution can be studied directly. This expansion was first made in \citeref{Berk1996PRL} in the case of a Krook collision operator, yielding the first iteration of the paradigmatic Berk-Breizman cubic (BBC) equation for the nonlinear evolution of the amplitude of marginally unstable waves. In that work, it was also demonstrated that the mode saturates at a level $\omegab = 8^{1/4}(1 - \gdgl)^{1/4}\nu_K$ such that the ordering $\omegabsq \ll \nu^2$ was guaranteed to be satisfied over the entire evolution, so long as the system was sufficiently close to marginal stability. Later this procedure was generalized to electromagnetic fluctuations in tokamak geometry\cite{Berk1997PPR,Berk1997report,Breizman1997POP} and also extended to include diffusive scattering and drag. In our case of interest for scattering and drag, the leading order expansion of \eqref{eq:vlasov} in $\omegabsq/\nu^2 \ll 1$ yields the following BBC equation:\cite{Lilley2009PRL} 

\begin{multline}
\deriv{A(\tau)}{\tau} = A(\tau) - \frac{1}{2}\int_0^{\tau/2}dz z^2 A(\tau-z) \int_0^{\tau-2z}dx \\
e^{-\nuhat^3 z^2 (2z/3 + x) + i \alphahat^2 z(z + x)}A(\tau-z-x)A^*(\tau-2z-x) 
\label{eq:cubic}
\end{multline}

For convenience, new dimensionless variables have been introduced which will be used for the rest of the analysis. The complex mode amplitude is normalized such that $A(\tau) \defined (\omegabsq/\gamma^2)/\sqrt{1 - \gdgl}$. All time scales are normalized to the net growth rate of the mode at $t = 0$, $\gamma = \gammal - \gammad$, such that $\tau \defined t\gamma$, $\nuhat = \nu/\gamma$, and $\alphahat = \alpha/\gamma$. 

The BBC equation contains rich nonlinear phenomena, and has previously been applied to interpret a variety of experimental phenomena including oscillations near the saturation level,\cite{Wong1997POP} pitchfork splitting,\cite{Fasoli1998PRL} chaotic mode evolution,\cite{Heeter2000PRL} and the emergence of frequency chirping.\cite{Duarte2017NF} The character of its solutions depend on the specific collisional rates $\nuhat$ and $\alphahat$. In particular, it contains both asymptotically steady state and dynamical (non-steady) solutions in the long time limit. While the collisional coefficients $\nu$ and $\alpha$ are treated as constants in this work for simplicity, in actuality they depend on phase space, so an integration over 6D phase space should be added to \eqref{eq:cubic}, as done in \citeref{Duarte2017NF}. Although the role of drag and scattering in determining the character of nonlinear behavior has been previously studied at some length, the modification of the quasi-steady solutions by drag has not been previously investigated in depth. Hence we focus on the regime of quasi-steady solutions, which occurs for $\rrat < 0.96$ and $\nuhat > \nuhat_\text{crit} \approx 2.05$. A diagram of the boundaries between different types of dynamical behavior in the $(\alphahat,\nuhat)$ parameter space is shown in Fig. 1 of \citeref{Lilley2009PRL}.  

Due to the enhanced effective scattering rates discussed earlier, the resonant particle dynamics can be strongly influenced by collisions, even in plasma conditions where the thermal particles are in a weakly collisional regime. Hence we seek a solution to \eqref{eq:cubic} when drag is subdominant to scattering and the effective scattering rate is much larger than the growth rate of the mode, \ie $\nuhat \gg 1$. This is similar to the case studied in \citeref{Duarte2019NF}, except generalized to include the effect of drag. This regime is common in tokamak plasmas, where previously examined DIII-D and NSTX discharges were often in the range of $\nuhat \gtrsim 10$ and $\alpha < \nu$.\cite{Duarte2017NF}

When $\nuhat \gg 1$, the magnitude of the integrand of \eqref{eq:cubic} is only nontrivial near $x = 0$ and $z \approx 1/\nuhat \ll 1$. Then the time-delayed mode amplitudes which appear, $A(\tau-z)$, $A(\tau-z-x)$, $A^*(\tau-2z-x)$, can each be approximated with argument $\tau$ only, allowing these terms to be pulled out of the integral. In other words, the form of the integral's kernel makes it so that only a small range of arguments for $A$ between $\tau$ and approximately $\tau - 3/\nuhat$ contribute to the integral. Within this range, we are assuming that the complex mode amplitude $A$ does not change much, such that it can be treated as a constant with respect to the integration variables. A rigorous justification of this approximation is given in \appref{app:rval}, and its consequences on the range of parameters ($\nuhat,\alphahat,\gdgl$) where it can be applied are discussed in \secref{sec:validity}. Normalizing the integration variables by $\nuhat$ transforms the integration limits to be $\nuhat\tau/2$ and $\nuhat\tau - 2\nuhat z$, both of which can be taken to infinity in the limit of large $\nuhat$ which we are examining. This allows the inner integral of \eqref{eq:cubic} to be performed exactly, leaving only the outer integral, which does not have a known solution. Hence, when $\nuhat$ is sufficiently large, one arrives at the much simpler time-local cubic (TLC) equation

\begin{align}
\label{eq:tlc}
\deriv{A(\tau)}{\tau} &= A(\tau) - b(\alphahat,\nuhat) A(\tau)\abs{A(\tau)}^2 \quad\text{where }\\ 
b(\alphahat,\nuhat) &= \frac{1}{2\nuhat^4}\int_0^\infty \frac{e^{-2 u^3/3 + i \alphahat^2 u^2/\nuhat^2}}{1 - i\alphahat^2 / (\nuhat^2 u)}du
\label{eq:bfull}
\end{align}

Hence we have shown that to leading order in $\omegabsq \ll \nu^2$ and $\gamma \ll \nu$, the complex mode amplitude obeys a nonlinear ODE which is much simpler than both the original PDE Vlasov system in \eqref{eq:vlasov} and the time-delayed integro-differential BBC equation in \eqref{eq:cubic}. Moreover, we note that \eqref{eq:tlc} is in the form of a Stuart-Landau equation,\cite{Landau1944,Stuart1958JFM} which appears in many contexts within nonlinear theory. All of the collisional physics is contained within the constant $b(\alphahat,\nuhat)$, which approximate expressions will be derived for in \secref{sec:dependence}. 

It is worthwhile to briefly summarize what brought us to this point. The assumption of $\omegabsq \ll \nu^2$ was used to derive the BBC equation (\eqref{eq:cubic}) from the Vlasov equation (\eqref{eq:vlasov}) and a power balance condition, yielding a time-delayed integro-differential equation for the evolution of the complex mode amplitude that is decoupled from the particle distribution $\df$. This first assumption destroys phase correlations within the resonant phase space island, but the evolution of the mode amplitude within the BBC equation still depends on the time history of the system. The additional assumption of $\gamma \ll \nu$ reduced the BBC equation to the TLC equation (\eqref{eq:tlc}), which is an ordinary differential equation where the system's evolution only depends on its current state. In other words, this second assumption also destroys time correlations during the growth of the mode. 

Whereas diffusive scattering is an irreversible process that destroys information, the drag component of the collision operator convects phase space structures coherently, tending to maintain the influence of the history on the future dynamics. Hence $\rrat$ is a key parameter in determining the validity of this approach. As will be discussed (and quantified) in \secref{sec:validity}, larger $\rrat$ requires larger $\nuhat$ in order for the time-local cubic equation to remain valid.  

\subsection{Solution of the time-local cubic equation: time evolution of the complex mode amplitude}
\label{sec:tlc-solution}

The TLC equation can be solved exactly, yielding an explicit analytic solution for the evolution of the complex mode amplitude. Analysis of the solution will demonstrate two important features: 1) the mode saturates at a larger amplitude when the ratio of drag to scattering $(\rrat)$ is increased, and 2) drag shifts the apparent frequency of oscillations by modulating the wave packet. To solve \eqref{eq:tlc}, decompose the complex mode amplitude into its time-varying magnitude and phase: $A(\tau) = \Aamp e^{i\phit}$ and also write the collisional integral $b$ explicitly in terms of its real and imaginary parts: $b(\alphahat,\nuhat) = \br + i \bi$. Then the real part of \eqref{eq:tlc} is 

\begin{align}
\Aamp' = \Aamp - \br\Aamp^3
\label{eq:aamptime}
\end{align}

This equation has the same form as Eq. 2 of \citeref{Duarte2019NF} (with $b \rightarrow \br$), so it has the same solution: 

\begin{align}
	\Aamp = \frac{\Aampz e^\tau}{\sqrt{1 - \br\Aampz^2(1 - e^{2\tau})}}
	\label{eq:aamp}
\end{align}

Above, $A_0 \defined A(0)$ is the initial amplitude, and $\abs{A_0}$ is its magnitude. Hence the saturation amplitude is 

\begin{align}
	\Asat \defined \lim_{\tau\rightarrow\infty}\Aamp = \frac{1}{\sqrt{\br}}
	\label{eq:asat}
\end{align} 

Since $\Asat$ is real by definition, this expression only makes sense when $\br > 0$. From numerical evaluation of \eqref{eq:bfull}, this condition exactly corresponds to $\rrat > 0.96$, which was the condition previously derived for the existence of steady state solutions.\cite{Lilley2009PRL} This expression for $\Asat$ first appeared in Lilley's PhD thesis,\cite{Lilley2009thesis} but the time-dependent solution was not previously derived.

The phase evolution can also be determined, as taking the imaginary part of \eqref{eq:tlc} yields 

\begin{align}
\phi'(\tau) = -\bi\Aamp^2
\label{eq:phitime}
\end{align}

Two important observations can be made here. First, the phase of the wave packet is constant unless $\bi$ is nonzero, which is the case if and only if the drag coefficient $\alpha$ is nonzero. Second, the phase will not evolve quickly until the wave approaches saturation since $\phi'$ is proportional to the square wave amplitude, which is small in the linear growth phase. \eqref{eq:phitime} can be integrated using \eqref{eq:aamp} to give 

\begin{align}
	\phit = \phiz - \frac{\bi}{2\br}\log\left[1 - \br\Aampz^2(1 - e^{2\tau})\right]
	\label{eq:phit}
\end{align}

Above, $\phiz \defined \phi(0)$ is the initial phase. At long times, the phase grows linearly in time, representing rotation at a constant rate in the complex plane: 

\begin{align}
	\frac{\domegasat}{\gamma} \defined -\lim_{\tau\rightarrow\infty}\phi'(\tau) = \frac{\bi}{\br}
	\label{eq:domegasat}
\end{align} 

Put together, the solution of \eqref{eq:cubic} in the large scattering $(\nuhat \gg 1)$ limit is given by 

\begin{align}
    A(\tau) = \frac{A_0 \exp\left\{\tau -i\frac{\bi}{2 \br}\log\left[1 - \br \Aampz^2 (1 - e^{2\tau})\right] \right\}}{\sqrt{1 - \br\Aampz^2(1 - e^{2\tau})}}
\label{eq:analytic}
\end{align}

The long time asymptotic behavior of the wave packet corresponds to regular oscillations at a fixed amplitude. 

\begin{align}
	\lim_{\tau\rightarrow\infty} A(\tau) = \Asat e^{-i\domegasat\tau/\gamma} = \frac{e^{-i(\bi/\br)\tau}}{\sqrt{\br}}
	\label{eq:asymptotic}
\end{align}

One can easily verify that \eqref{eq:asymptotic} is a solution to the BBC equation in the limit of $\tau \rightarrow \infty$. 

\begin{figure}[tb]
\includegraphics[width=\thirdwidth]{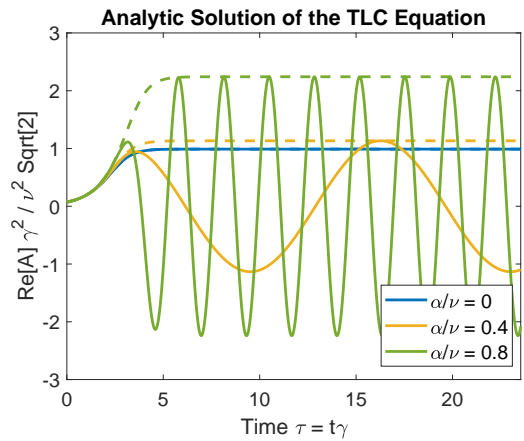}
\caption{Real part of the analytic solution of the time-local cubic equation for $\rrat = 0$, 0.4, 0.8 (\eqref{eq:analytic})). Dashed curves show the magnitude (including the imaginary part not shown).}
\label{fig:solution}
\end{figure}

The analytic evolution of the real part of the mode amplitude $A$ is shown in \figref{fig:solution} for $\rrat = 0$, 0.4, and 0.8. The dashed curves correspond to the magnitude $\abs{A}$. In each case, there is an initial linear phase of exponential growth where the nonlinearities are insignificant, and the mode amplitude essentially obeys the simple relation $dA/d\tau = A$. As the mode enters the nonlinear phase, the cubic term becomes relevant, leading to saturation of $\abs{A}$. In the absence of drag, $A$ does not oscillate since $\bi = 0$. The $\rrat = 0.4$ curve demonstrates that adding drag both introduces oscillations and increases the saturation level relative to the scattering only case. Both the saturation level and frequency of oscillations in $A$ increase substantially when $\rrat$ is increased to 0.8. The saturation level is more than double the scattering only case and nearly double the case of $\rrat = 0.4$. This is an important conclusion -- the effect of drag becomes much more pronounced as the ratio of drag to scattering approaches the steady/non-steady solution boundary at $\rrat = 0.96$. The dependence of the saturation level and frequency shift on $\rrat$ will be quantified in \secref{sec:dependence}. Some additional interesting properties of the analytic TLC solution are given in \appref{app:quantities}.

\subsection{Interpretation of the frequency shift due to drag}
\label{sec:fshift}

The interpretation of the derived frequency shift is somewhat subtle, so it is worth elaborating on its meaning and implications. The oscillations at $\domegasat$ are of the wave packet itself, which are different (and typically much slower) than the oscillations at the linear wave frequency $\omega$. However, the modulation of the wave packet does not result in beating between $\domegasat$ and $\omega$, but rather a finite shift in the apparent nonlinear oscillation frequency. To see this, recall how the total wave field is connected to the quantities we have just solved for. $E(x,t) = \Ehat e^{i(kx-\omega t)} \propto \Aamp e^{i\phi(\tau)}e^{i(kx-\omega t)} \xrightarrow{\tau\rightarrow\infty} \Asat e^{i(kx-(\omega + \domegasat)t)}$. This frequency shift is the direct result of including drag in the kinetic equation, as without it, $A(\tau)$ would be purely real (equivalently, $\bi = 0$ when $\alpha = 0$, so $\domegasat = 0$ as well). As mentioned earlier, larger values of $\rrat$ induce a larger frequency shift. 

\begin{figure}[tb]
\includegraphics[width=\thirdwidth]{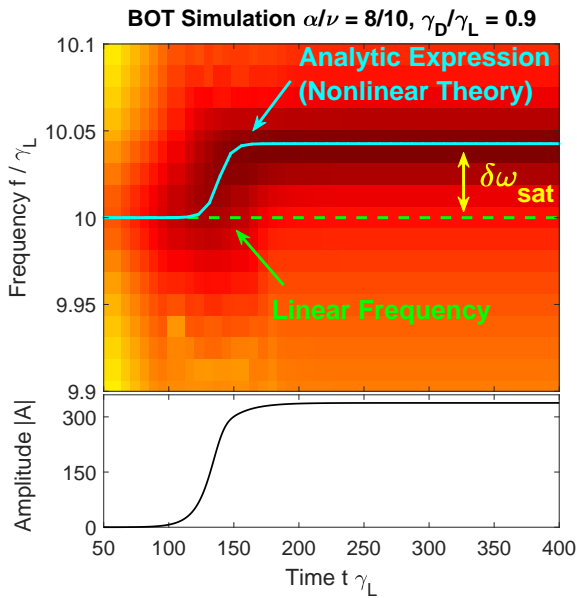} 
\caption{Top: spectrogram of \BOT simulation with $\nuhat = 10$, $\alphahat = 8$, $\gammad/\gammal = 0.9$. The linear oscillation frequency is shown as a dashed green line. The blue curve shows the time-dependent nonlinear frequency predicted by theory (derivative of \eqref{eq:phit}). Bottom: magnitude of the mode amplitude from same simulation.}
\label{fig:specgram}
\end{figure}

The frequency shift is a nonlinear effect -- it is not present until the mode nears saturation. The transition from the linear frequency to the nonlinear one (including the shift) can be visualized in a spectrogram of the total fluctuation: $\Re{E(x,t)}$. In the top of \figref{fig:specgram}, a sliding Fourier transform of the fluctuation is shown from a 1D Vlasov simulation using the \BOT code.\cite{Lilley2010POP} The bottom panel shows the evolution of the magnitude of the mode amplitude $\abs{A}$. Further quantitative comparisons with this code will be presented in \secref{sec:bot}. For the simulation shown in \figref{fig:specgram}, $\nuhat = 10$, $\alphahat = 8$, and $\gdgl = 0.9$ were used as input parameters. The eigenmode was set to have a linear oscillation frequency of $f = 10\gammal$, shown by the green dashed line. Early in the simulation, \eg during the linear growth phase, the spectrogram is peaked around this frequency. As the mode enters the nonlinear phase, the peak in the spectrogram shifts upwards before reaching a new oscillation frequency, which it maintains for the rest of the simulation. The difference between the final nonlinear frequency and the initial linear frequency is the frequency shift due to drag. The time evolution of the frequency in the simulation closely tracks the time-dependent frequency derived in \eqref{eq:phitime}, overlaid with the blue curve. 

The frequency shift does not appear to be large relative to the mode frequency. The simulation was run with $\rrat = 0.8$ in order to show a relatively large frequency shift that would not be obscured by the resolution of the spectrogram. Smaller values of $\rrat$ would have even smaller shifts, and $\rrat$ can not be increased above 0.96 without changing the nature of the solution from a steady state to a non-steady/dynamical one. Even so, the difference between the linear and nonlinearly-shifted frequencies is very small -- about $0.5\%$. Partially this is because $f = 10\gammal$ was chosen arbitrarily, which the frequency shift does not depend on. A more meaningful measure is how the frequency shift relates to the growth rate. As derived, the frequency shift for $\rrat = 0.8$ is approximately 2$\gamma$ (see \figref{fig:fsat} in the next section). Meanwhile, $\gamma \ll \omega$ is usually satisfied for weakly driven modes, implying that $\domegasat \ll \omega$ is likely as well, unless $\alpha/\nu$ is very close to 0.96, where the frequency shift becomes very large. Hence this frequency shift probably does not have to be considered when interpreting experimental spectrograms under ordinary conditions. 

Even if the size of the shift is not substantial, the fact that the frequency changes as the mode evolves has implications for the resonant wave-particle interaction. As derived in \citeref{Duarte2021U}, drag also modifies the resonance condition by shifting the velocity where the strongest wave-particle interaction takes place. Hence, the resonant velocity is given by 

\begin{align}
	\omega(t) - k \vres(t) = \Delta\Omega \approx \frac{\Gamma[4/3]\alpha^2}{2\nu}
	\label{eq:res-shift}
\end{align}

However, since the frequency changes due to drag as the mode evolves, the velocity of the most strongly interacting particles must also change in order to continuously satisfy the resonance condition. In other words, the nonlinear frequency shift $\domegasat$ induces a corresponding shift $\dvres$ in the region of velocity space where the resonant wave particle interaction is strongest. Initially in the linear stage of the instability, there is a resonant velocity $\vresl$ satisfying 

\begin{align}
	\omegal - k\vresl = \Delta\Omega
\end{align}

Then in the nonlinear phase once the frequency has shifted, the resonant velocity also shifts to $\vresnl = \vresl + \dvres$ where $\dvres = \domegasat/k$ such that the resonance condition remains satisfied as the mode grows. 

\begin{align}
	\omegal + \domegasat - k\left(\vresl + \dvres\right) = \Delta\Omega
\end{align}

In practice $\dvres$ is small since $\domegasat$ is small. Moreover, the resonance condition only describes the velocity of strongest wave-particle interaction -- it does not need to be satisfied exactly in order for a net energy change to occur except in the absence of collisions (and also neglecting finite amplitude effects). In reality, the resonance is broadened by collisions such that particles with velocities in the range $\Delta\Omega - \nu \lesssim \omega - k v \lesssim \Delta\Omega + \nu$ can efficiently interact with the wave. Hence the nonlinear modification of the resonance due to drag is more of fundamental than practical interest. 

\subsection{Dependence of the saturation amplitude and frequency shift on drag and scattering} 
\label{sec:dependence}

To this point, we have derived an analytic expression for $A(\tau)$ and also its long time asymptotic behavior. These solutions (in \eqref{eq:analytic} and \eqref{eq:asymptotic}) depend on the the real and imaginary parts of the integral function $b(\alphahat,\nuhat)$ defined in \eqref{eq:bfull}. In order to understand the dependence of the saturation amplitude and nonlinear frequency shift on the ratio of drag to scattering, we must calculate $\br$ and $\bi$. The real and imaginary parts of $b$ can be rewritten from \eqref{eq:bfull} as

\begin{multline}
\label{eq:brint}
\br = \frac{1}{2\nuhat^4}\int_0^\infty \frac{e^{-2u^3/3}}{1 + \alpha^4/(\nu^4 u^2)} \times \\ 
\left[\cos\left(\frac{\alpha^2 u^2}{\nu^2}\right) - \frac{\alpha^2 }{\nu^2 u}\sin\left(\frac{\alpha^2 u^2}{\nu^2}\right)\right]du 
\end{multline} 

\begin{multline}
\bi = \frac{1}{2\nuhat^4}\int_0^\infty \frac{e^{-2u^3/3}}{1 + \alpha^4/(\nu^4 u^2)} \times \\
\left[\sin\left(\frac{\alpha^2 u^2}{\nu^2}\right) + \frac{\alpha^2}{\nu^2 u} \cos\left(\frac{\alpha^2u^2}{\nu^2}\right)\right]du
\label{eq:biint}
\end{multline}

The above integrals are not known in terms of elementary or special functions, so in general they must be evaluated numerically. However, approximate analytic expressions can also be derived in the limit of $\alpha \ll \nu$, which is a typical (although not universal) condition satisfied in modern tokamaks. Due to the decaying cubic exponential, the integrands become vanishingly small for $u > 2$. Hence, $\alpha \ll \nu$ justifies $\alpha^2 u^2/\nu^2 \ll 1$ expansions of the trigonometric functions as well. 
To lowest order in $\rrat$, the following is found: 


\begin{align*}
\label{eq:brappx}
\br &\stackrel{\alpha \ll \nu}{\approx} \frac{1}{2\nuhat^4}\int_0^\infty \frac{e^{-2u^3/3}}{1 + \alpha^4/(\nu^4u^2)}du \numberthis \\ 
\label{eq:brmg}
&= \frac{1}{16\pi^{5/2}\nuhat^4}\frac{\alpha^2}{\nu^2}\MGsimp{5}{3}{-\frac{1}{6},\frac{1}{6},\frac{1}{2}}{-\frac{1}{6},0,\frac{1}{6},\frac{1}{2},\frac{1}{2}}{\frac{\alpha^{12}}{9\nu^{12}}} \numberthis \\
\label{eq:biappx}
\bi &\stackrel{\alpha \ll \nu}{\approx} \frac{1}{2\nuhat^4}\frac{\alpha^2}{\nu^2}\int_0^\infty \frac{e^{-2u^3/3}}{1 + \alpha^4/(\nu^4u^2)}\left[u^2 + \frac{1}{u}\right]du \numberthis \\ 
\label{eq:bimg}
&= \frac{1}{16\pi^{5/2}\nuhat^4} \frac{\alpha^2}{\nu^2}\left[\frac{\alpha^6}{\nu^6} \MGsimp{5}{3}{-\frac{1}{2},-\frac{1}{6},\frac{1}{6}}{-\frac{1}{2},-\frac{1}{6},0,\frac{1}{6},\frac{1}{2}}{\frac{\alpha^{12}}{9\nu^{12}}} \right. \\ 
& \left. + \MGsimp{5}{3}{0,\frac{1}{3},\frac{2}{3}}{0,0,\frac{1}{3},\frac{1}{2},\frac{2}{3}}{\frac{\alpha^{12}}{9\nu^{12}}}\right] \numberthis
\end{align*}

After approximating the integrands to lowest order in the small parameter $\alpha^2 u^2/\nu^2 \ll 1$, the integration is performed exactly using the very general Meijer $G$-function\cite{Meijer1941,NISTmath} which is represented in \Mathematica as

\begin{multline}
\MG{m}{n}{p}{q}{a_1, \dots, a_n}{a_{n+1},\dots, a_p}{b_1, \dots, b_m}{b_{m+1}, \dots, b_q}{z} \defined \\ 
\text{MeijerG}\left[\left\{\left\{a_1,\dots,a_n\right\},\left\{a_{n+1},\dots,a_p\right\}\right\}, 
\right. \\ \left.
\left\{\left\{b_1,\dots,b_m\right\},\left\{b_{m+1},\dots,b_q\right\}\right\},z\right]
\end{multline}

The expressions contained in \eqref{eq:brmg} and \eqref{eq:bimg} provide no more insight than \eqref{eq:brint} and \eqref{eq:biint}. However, they are useful in that their series expansion can be computed analytically in order to develop more transparent approximations. Thus when drag is much smaller than scattering $(\alpha \ll \nu)$, we find 


\begin{align}
\label{eq:brappxsmall}
    \br &\stackrel{\alpha \ll \nu}{\approx} \frac{1}{2\nuhat^4}\left[ (1/3)\Gamma(1/3)(3/2)^{1/3} - \frac{\pi}{2}\frac{\alpha^2}{\nu^2}\right] \plusord{\frac{\alpha^4}{\nu^4}} \\
\label{eq:biappxsmall} 
	\bi &\stackrel{\alpha \ll \nu}{\approx} \frac{1}{6\nuhat^4}\left[\frac{3}{2} -\eulergamma + \log\frac{3}{2} - 6\log\frac{\alpha}{\nu}\right]\frac{\alpha^2}{\nu^2} \plusord{\frac{\alpha^6}{\nu^6}}
\end{align}

Above, $\eulergamma$ is the the Euler-Mascheroni constant, defined as $\eulergamma = \lim_{n\rightarrow\infty} \left(-\log{n} + \sum_{k=1}^\infty \frac{1}{k}\right) \approx 0.577$. The leading order approximation for $\bi$ is accurate to within $9\%$ for all $0 < \rrat < 1$, with error less than $1\%$ for $\rrat < 0.5$. In contrast, the leading order approximation for $\br$ has a maximum relative error of $5\%$ for $\rrat < 0.5$, becoming poor for larger $\rrat$. Higher order expansions of the trigonometric functions could have been used in \eqref{eq:brappx} and \eqref{eq:biappx} for improved accuracy, which would result in additional Meijer $G$-functions in the resulting expressions. However, these terms ultimately do not modify the leading order series expansions given in \eqref{eq:brmg} and \eqref{eq:bimg}, so we chose to neglect them at that intermediate step. The utility of these approximations for $\br$ and $\bi$ is in giving a sense of how the saturation amplitude and frequency shift depend on $\rrat$, at least when this ratio is relatively small. To that end, we can use these expressions to write 

\begin{figure}[tb]
\subfloat[\label{fig:asat}]{\includegraphics[width = \thirdwidth]{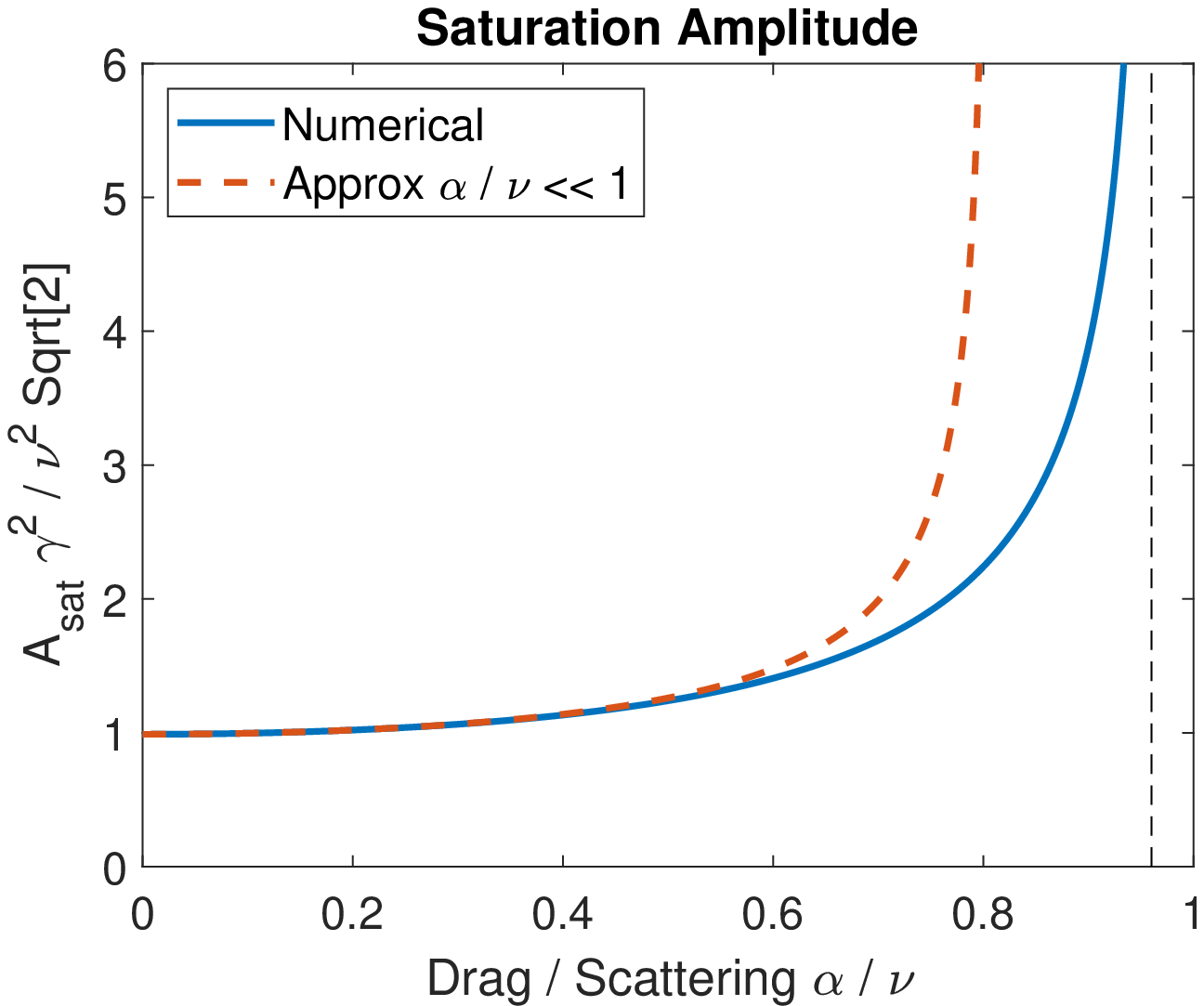}} \\
\subfloat[\label{fig:fsat}]{\includegraphics[width = \thirdwidth]{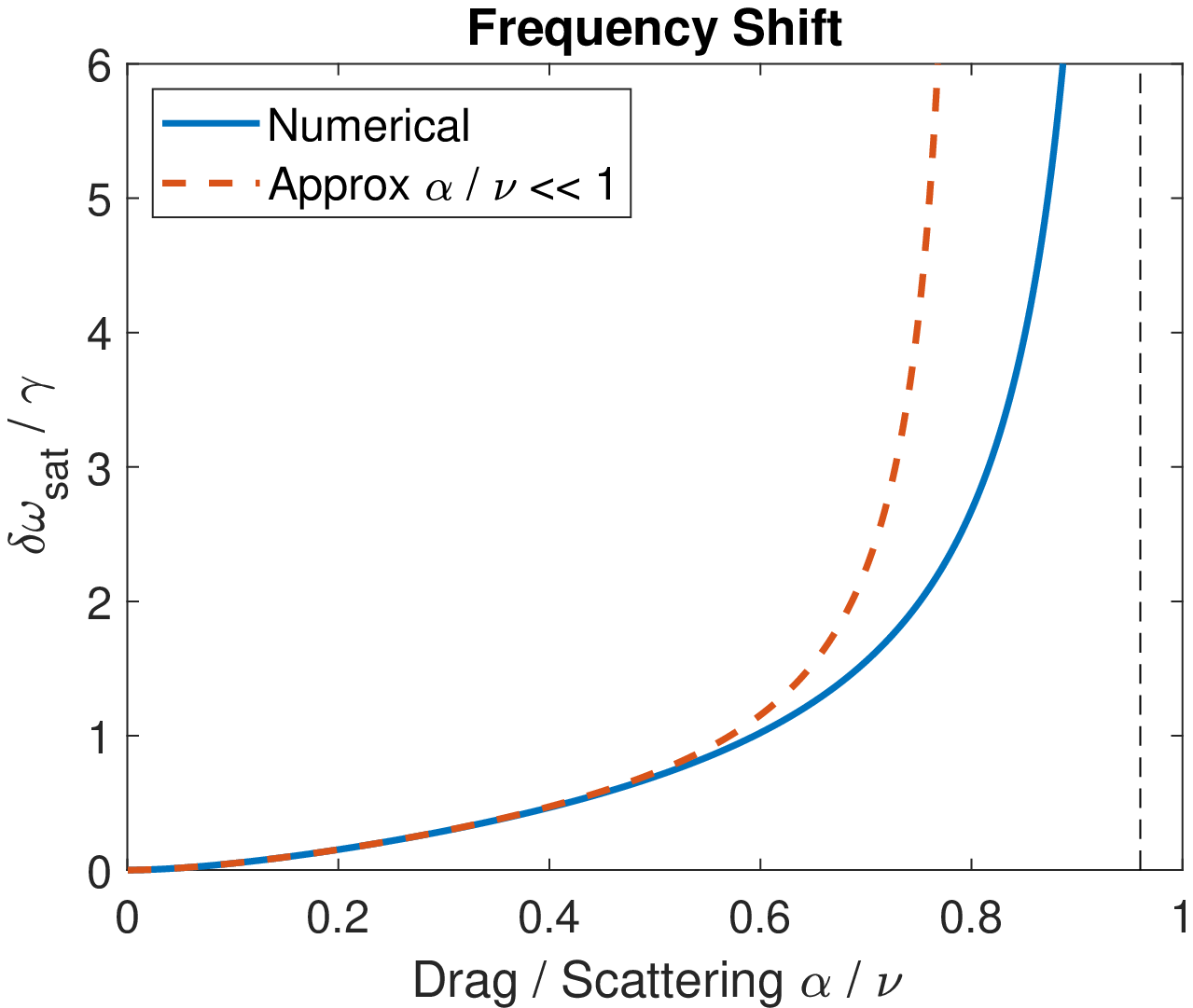}}
\caption{Dependence of (a) saturation amplitude (\eqref{eq:asat}) and (b) frequency shift (\eqref{eq:domegasat}) on the ratio of drag to scattering. Solid curves use numerically integrated values, while the dashed curves are analytic approximations valid for $\rrat \ll 1$ (\eqref{eq:asat-approx} and \eqref{eq:fsat-approx}). Vertical black dashed line indicates the boundary between steady and non-steady solutions.}
\label{fig:sat}
\end{figure}


\begin{align}
\label{eq:asat-approx}
\Asat &= \frac{1}{\sqrt{\br}} \approx \frac{\nuhat^2\sqrt{2}}{\sqrt{(1/3)\Gamma(1/3)(3/2)^{1/3} - \pi\alpha^2/(2\nu^2)}} \\ 
\label{eq:fsat-approx}
\domegasat &= \gamma\frac{\bi}{\br} \approx 
\gamma\frac{\left[\frac{3}{2} -\eulergamma + \log\frac{3}{2} - 6\log\frac{\alpha}{\nu}\right]}
{\Gamma(1/3)(3/2)^{1/3} - 3\pi\alpha^2/(2\nu^2)}\frac{\alpha^2}{\nu^2}
\end{align}

Setting $\alpha = 0$ in \eqref{eq:asat-approx} recovers the previously derived formula for saturation amplitude due to scattering only:\cite{Duarte2019NF} $\Asat = 1.4\nuhat^2$, or in terms of the bounce frequency of earlier works,\cite{Berk1997PPR,Petviachvili1999thesis} $\omegab = 1.18 \left(1 - \gammad/\gammal\right)^{1/4}\nu$. Similarly, evaluating \eqref{eq:fsat-approx} at $\alpha = 0$ removes the frequency shift entirely, as expected. In terms of un-normalized experimental quantities, the saturation amplitude including drag then is 

\begin{align}
\delta E_\text{sat} \approx \frac{m_i\nu^2}{q_i k}\sqrt{\frac{2\left(1 - \gammad/\gammal\right)}{(1/3)\Gamma(1/3)(3/2)^{1/3} - \pi \alpha^2/2\nu^2}}
\label{eq:asat-E}
\end{align}

It is worth pointing out that while both the saturation amplitude and frequency shift are increasing functions of $\rrat$, the saturation amplitude scales with the square of the effective scattering rate $\Asat \propto \nuhat^2 = \nu^2/\gamma^2$ while the frequency shift instead scales with the net growth rate of the mode $\gamma$. These are general trends that do not depend on the approximation $\alpha \ll \nu$ used to derive \eqref{eq:asat-approx} and \eqref{eq:fsat-approx}. Larger scattering rate leads to larger growth rate because the collisions are responsible for replenishing the driving gradient of $\fz$ which is otherwise flattened as the wave grows. The exact dependence of $\Asat$ and $\domegasat$ on $\rrat$ are found by numerical integration and compared to our approximate analytic expressions in \figref{fig:sat}. As expected, the approximate expressions are very accurate for $\rrat < 0.5$ but lose accuracy for larger values of this ratio, inheriting the error in the approximation for $\br$. The modification of the saturation amplitude and frequency shift relative to the case of no drag $(\alpha = 0)$ is most substantial when drag becomes comparable to scattering. As will be discussed in \secref{sec:bot}, these trends compare well with 1D Vlasov simulations. Clearly both quantities diverge at the steady/non-steady solution boundary of $\rrat = 0.96$, marked by a vertical black dashed line on \figref{fig:sat}. This divergence is not physical, but rather a breakdown of the theoretical formalism, indicating that higher order nonlinearities would become important. With these newly derived quantities, the validity of the TLC equation can now be addressed more precisely. 

\subsection{Validity of the time-local cubic equation}
\label{sec:validity}

Deriving the TLC equation, and its subsequent solution, required two main assumptions to be valid. Knowledge of the saturation level and wave packet oscillation frequency allow us to translate these assumptions into quantitative constraints. It will now be shown that as $\rrat$ is increased, these assumptions become more restrictive such that 1) the system must be closer to marginal stability, and 2) the ratio of scattering to net growth rate must increase. 

First, $\omegabsq \ll \nu^2$ was used to derive the BBC equation from the Vlasov equation. Since $\omegab^2 \like A$, ensuring that $\omegabsq \ll \nu^2$ is satisfied at saturation will ensure that it is obeyed at all times. This ends up placing a constraint on how close the system must be to marginal stability. To see this, define a new function $I^{-2}(\rrat) = \nuhat^4 \brvar$. This function is useful because it isolates the dependence of the saturation level on $\rrat$ from its additional dependence on $\nu$. Then rewriting the normalized variables in terms of the bounce frequency, we find that $\Asat = 1/\sqrt{\br}$ is equivalent to $\abs{\omegabsat^2}/\nu^2 = I(\rrat)\sqrt{1 - \gammad/\gammal}$. In order for $\abs{\omegabsat^2}/\nu^2 \ll 1$, as assumed at the outset, we must require that 

\begin{align} 
1 - \gammad/\gammal \ll \frac{1}{I^2(\rrat)}
\label{eq:gdgllim} 
\end{align} 

\begin{figure}[tb]
\subfloat[\label{fig:gdglbound}]{\includegraphics[width = \thirdwidth]{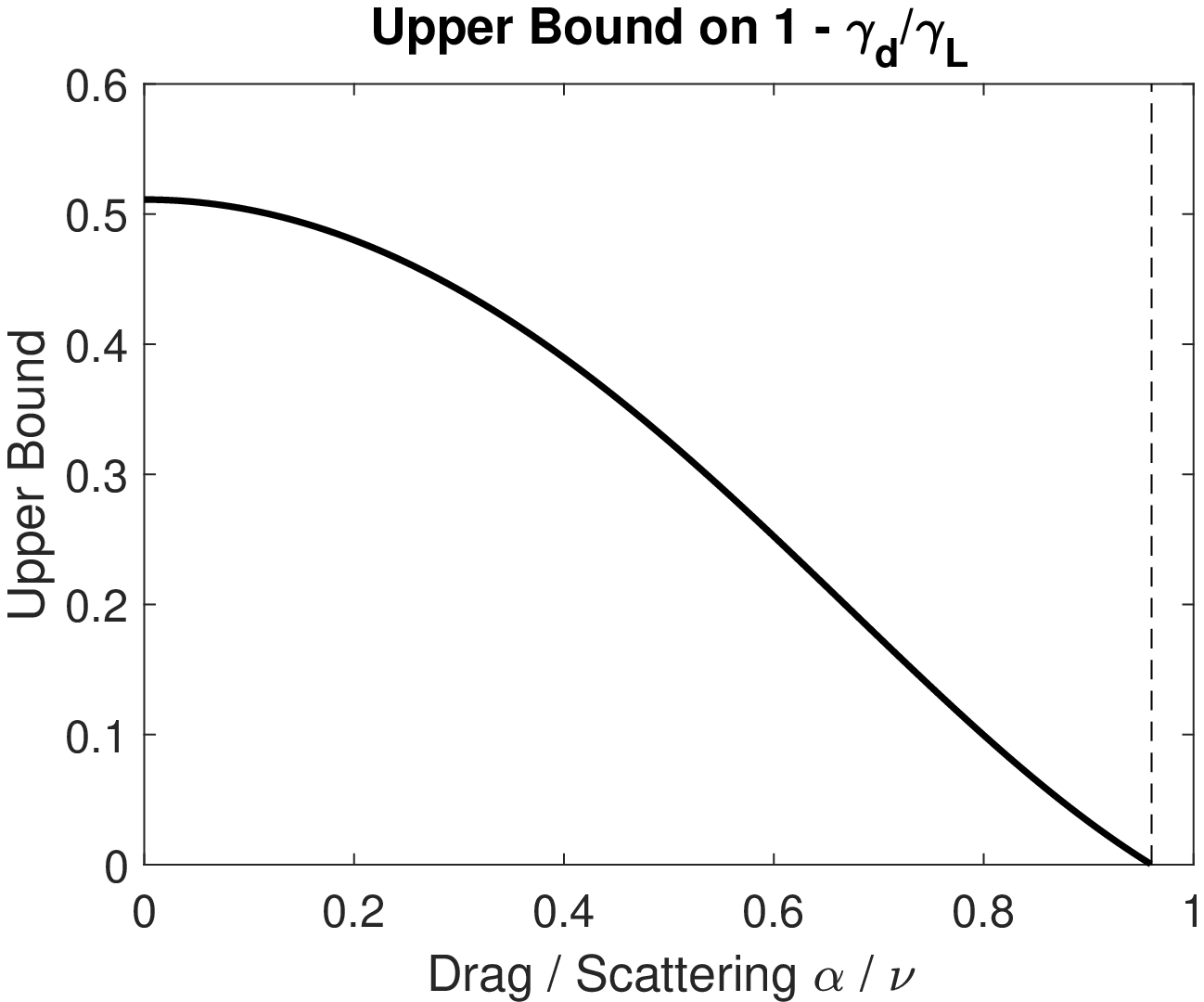}} \\
\subfloat[\label{fig:nubound}]{\includegraphics[width = \thirdwidth]{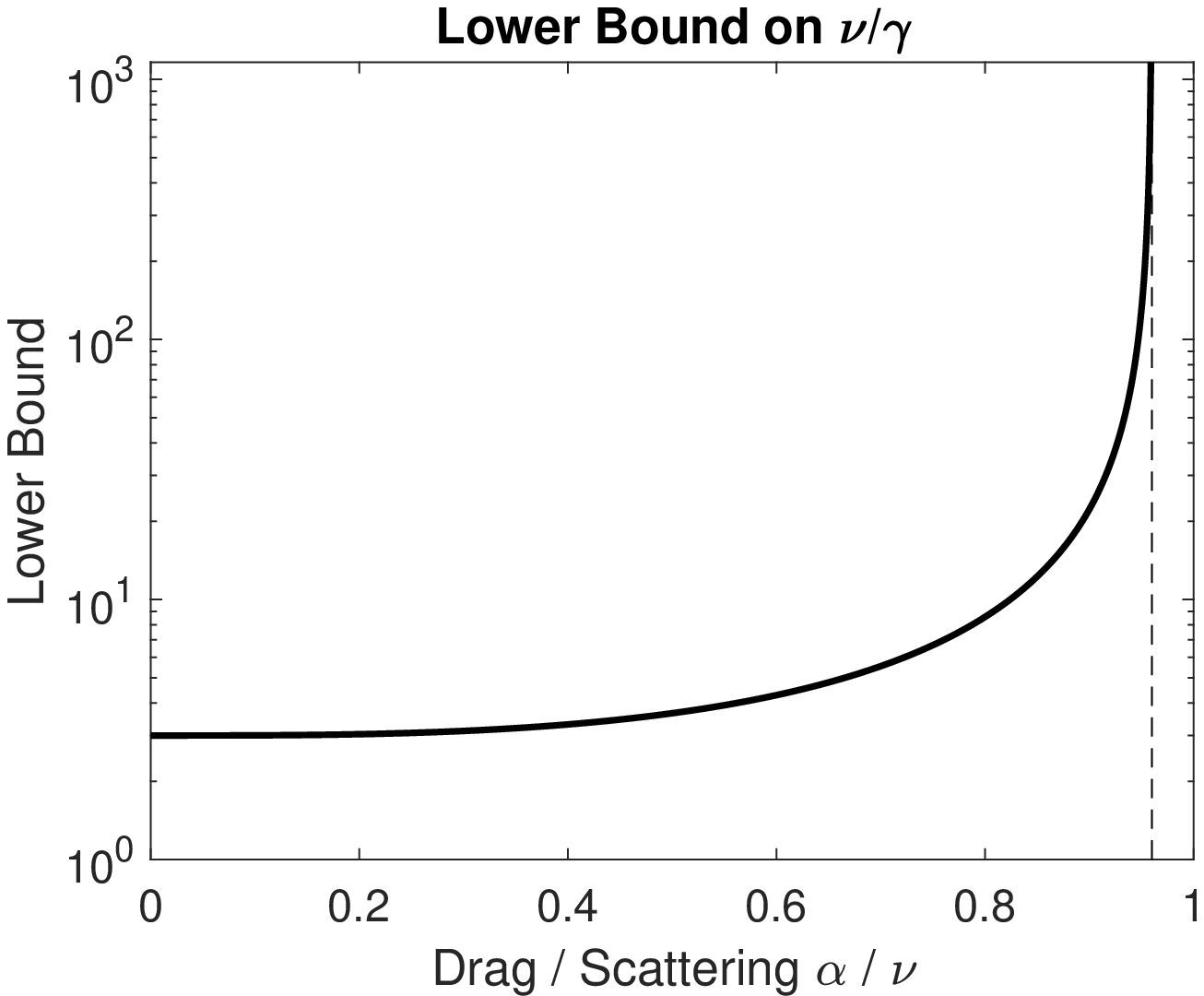}}
\caption{Constraints on system parameters for applicability of the TLC equation as a function of $\rrat$. (a) Upper bound on $1 - \gdgl$ (departure from marginal stability, \eqref{eq:gdgllim}), linear scale and (b) Lower bound on $\nu/\gamma$ (collisional dominance over growth, \eqref{eq:nuhatlim}), log scale.}
\label{fig:bounds}
\end{figure}

The right hand side of this equation is plotted in \figref{fig:gdglbound}. In the case of no drag ($\alpha = 0$), this constraint simply becomes $1 - \gdgl \ll 0.51$, which is well satisfied by systems that are typically considered close to marginal stability. However, when $\rrat$ approaches the steady/non-steady solution boundary of $\rrat = 0.96$, the function $I^{-2}(\rrat)$ becomes very small, requiring the system to be closer to marginal stability. As some reference points, $\rrat = 0.5$ requires $1 - \gdgl \ll 0.33$, while $\rrat = 0.9$ requires $1 - \gdgl \ll 0.03$. For any value of $\rrat < 0.96$, there exists a sufficiently small value of $1 - \gdgl$ such that $\omegabsq \ll \nu^2$ will be satisfied as required, but the required closeness to marginal stability becomes increasingly restrictive for large values of $\rrat$. 

Second, the ``time-local'' approximation was used to ignore the time delays in the BBC equation and arrive at the TLC equation. The reasoning here was that the range of time delays contributing to the integral vanished as $\nuhat = \nu/\gamma \rightarrow \infty$, but rigorously quantifying how large $\nuhat$ must be to justify this procedure requires more subtle arguments. A detailed derivation is given in \appref{app:rval}, with the main concepts outlined here. 

Essentially, ignoring the time delays is justified when $A$ does not change very much between $\tau$ and $\tau - \delta$, where $\delta$ is the largest time delay appearing in \eqref{eq:cubic}. This condition is equivalent to requiring that $\abs{A'(\tau)}/\abs{A(\tau)} \ll 1$. The complex mode amplitude $A$ has two components which vary on different timescales. First, the magnitude $\abs{A(\tau)}$ varies no faster than the initial growth rate $\gamma = \gammal - \gammad$ since the net growth rate of the mode decreases as the distribution is flattened, eventually going to zero at saturation. Second, the phase $\phi(\tau)$ varies no faster than the frequency shift $\domegasat$. The combined variation in magnitude and phase must obey $\abs{A'(\tau)}/\abs{A(\tau)} \ll 1$ to justify the time-local approximation, which yields the following constraint on $\nuhat = \nu/\gamma$

\begin{align}
		\nuhat = \frac{\nu}{\gamma} \gg 3\sqrt{1 + \frac{\bi^2}{\br^2}}
		\label{eq:nuhatlim}
\end{align}

Here the first term under the radical bounds the rate of change of the magnitude and the second term bounds the rate of change of the phase. The right hand side of this equation is plotted in \figref{fig:nubound}. Without any drag, the condition is $\nuhat \gg 3$, explaining the good agreement found between the analytic solution and cubic equation in the absence of drag in \citeref{Duarte2019NF}. For small values of drag, the constraint does not change much since $\domegasat$ remains small. At $\rrat \approx 0.6$, the two terms under the radical equal one another and the condition becomes $\nuhat \gg 4.2$. As $\rrat$ is increased further, the constraint becomes much more restrictive due to the rapid oscillation of the wave packet when drag is comparable to scattering. When $\rrat = 0.9$, the constraint increases to $\nuhat \gg 22$. Nonetheless, $\nuhat$ can always be made sufficiently large to satisfy \eqref{eq:nuhatlim} for any value of $\rrat$ which admits a steady solution. 

To summarize, for a given ratio of drag to scattering $\rrat$, \eqref{eq:gdgllim} specifies how small $1 - \gdgl$ must be to ensure that the BBC equation is valid. Additionally, \eqref{eq:nuhatlim} specifies how large $\nuhat = \nu/\gamma$ must be be in order to ensure that the TLC equation (\eqref{eq:tlc}) and its subsequent solution will be valid for the entire evolution of the mode. If only the constraint on $1 - \gdgl$ is satisfied but not the restriction on $\nuhat$, then the expressions for the final saturation level (\eqref{eq:asat}) and frequency shift (\eqref{eq:domegasat}) will still be valid (since they can be found by taking the $\tau\rightarrow\infty$ limit of the BBC equation), but the TLC solution specifying the amplitude at all times (\eqref{eq:analytic}) will not be. 
In particular, note that in the idealized limit of marginal stability where $\gamma \rightarrow 0$, both of the constraints are automatically satisfied for all values of $\rrat$.  

\newcommand{\figfourlen}{0.33\textwidth}
\begin{figure*}[tb]
\subfloat[\label{fig:nuvary-3}]{\includegraphics[width = \figfourlen]{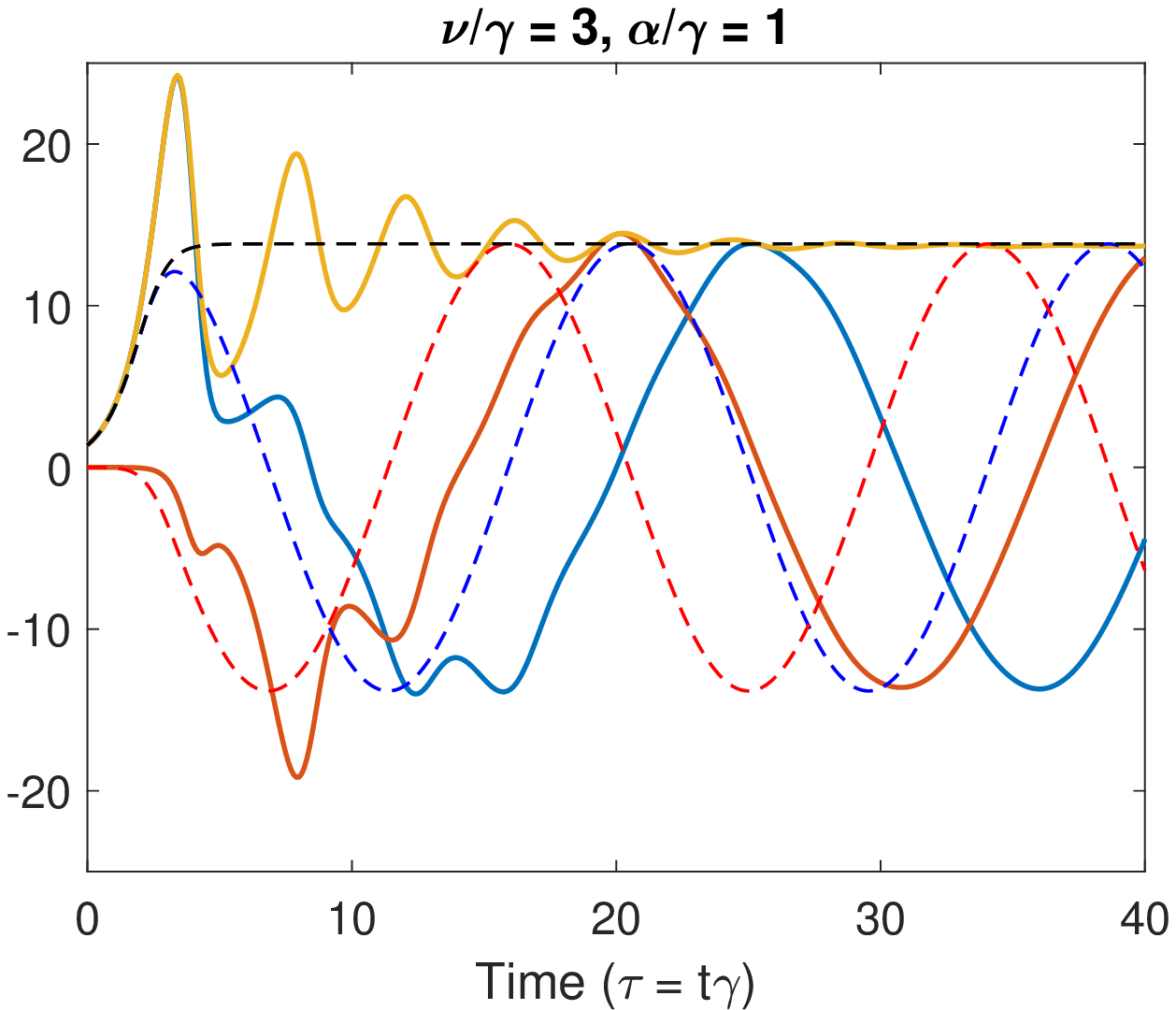}}
\subfloat[\label{fig:nuvary-10}]{\includegraphics[width = \figfourlen]{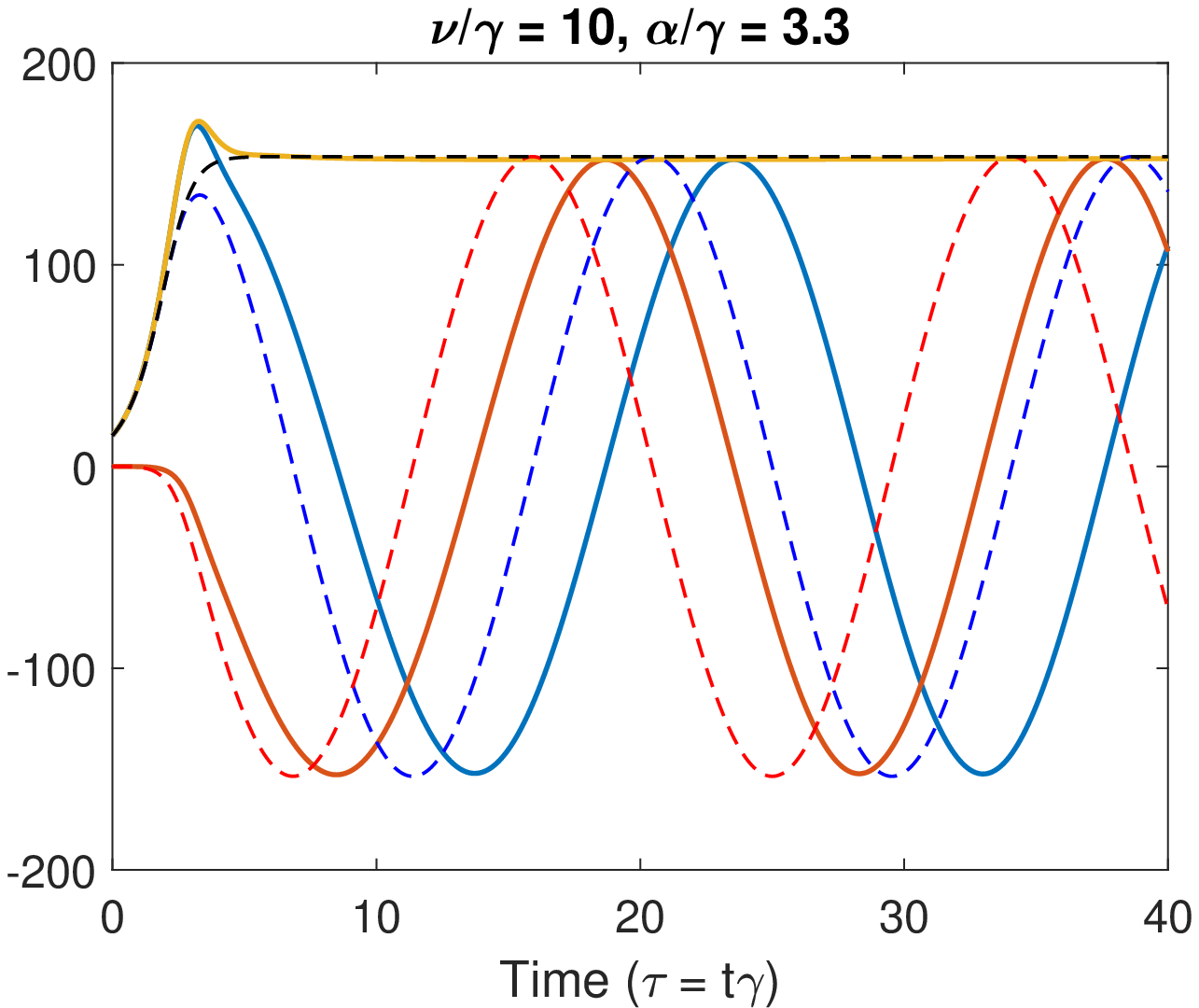}} 
\subfloat[\label{fig:nuvary-30}]{\includegraphics[width = \figfourlen]{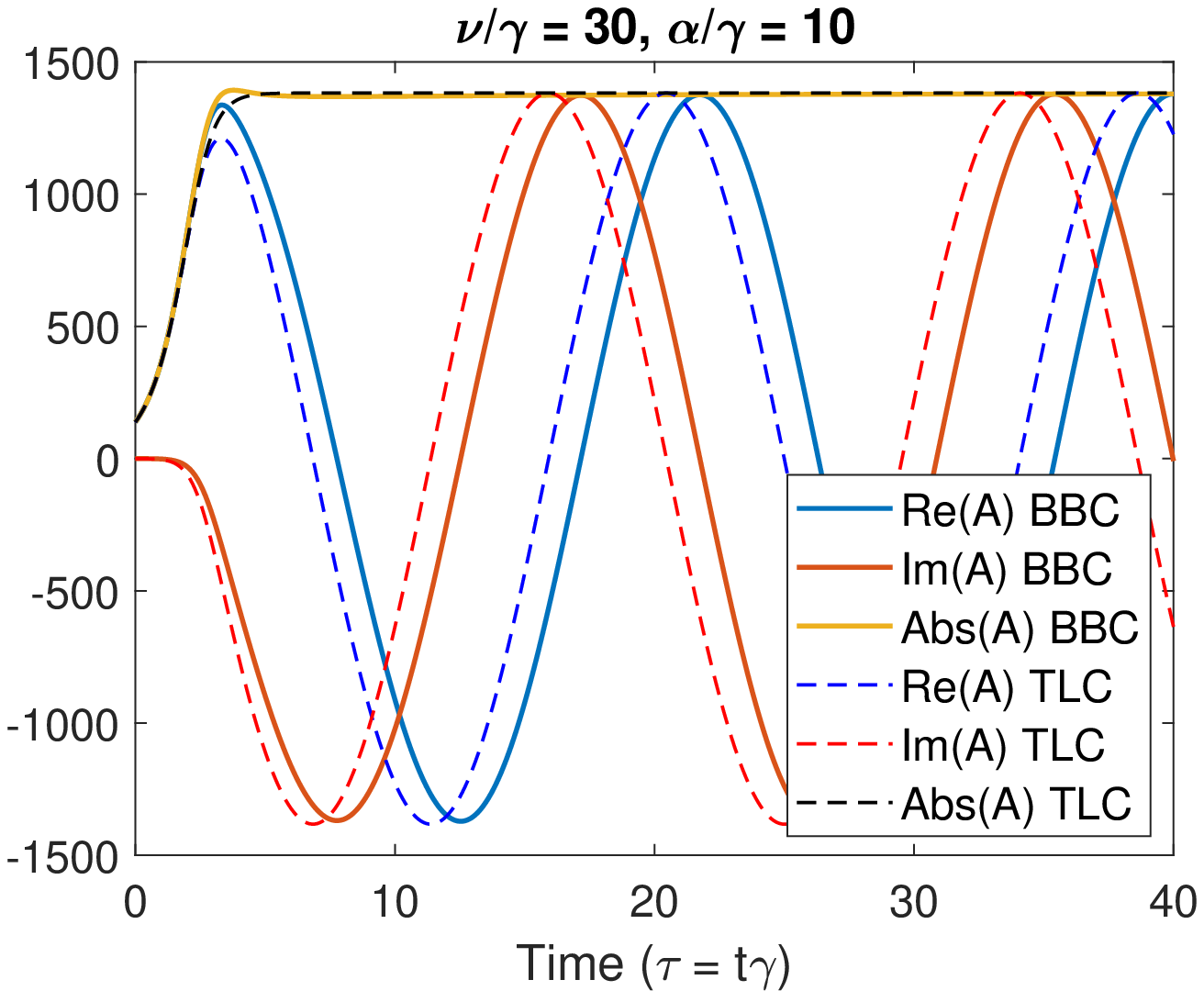}} \\
\subfloat[\label{fig:avary-1}]{\includegraphics[width = \figfourlen]{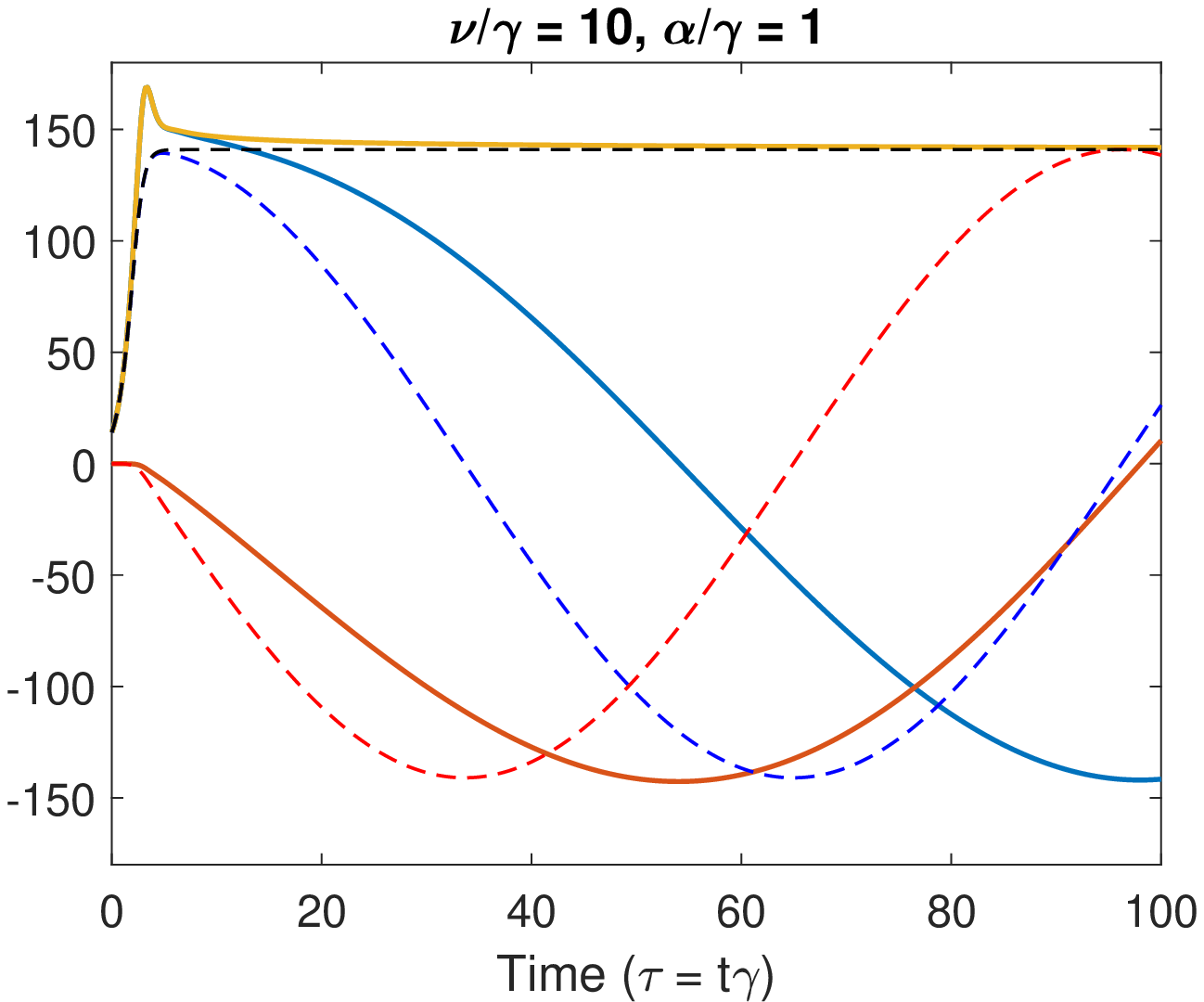}}
\subfloat[\label{fig:avary-2}]{\includegraphics[width = \figfourlen]{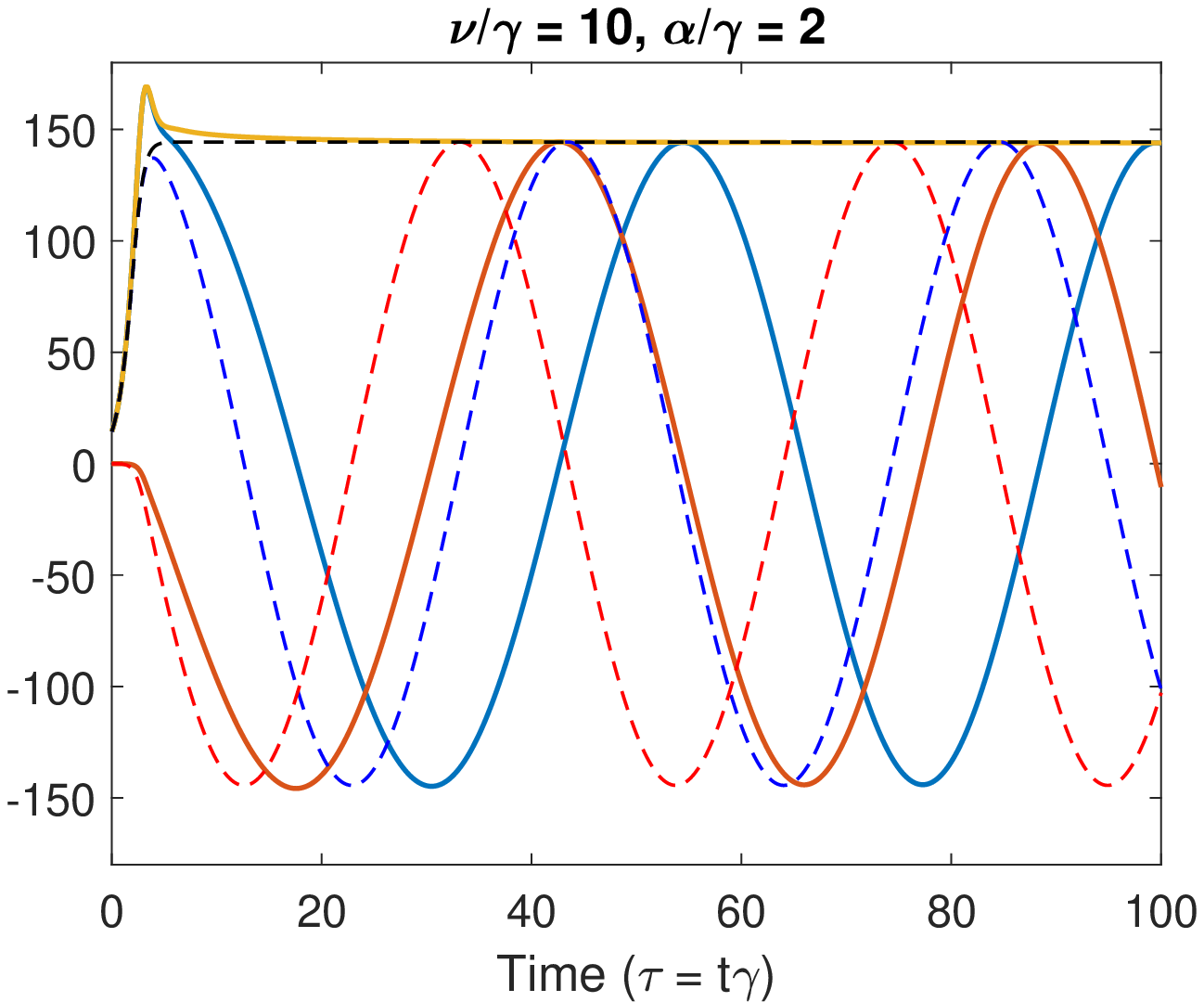}} 
\subfloat[\label{fig:avary-4}]{\includegraphics[width = \figfourlen]{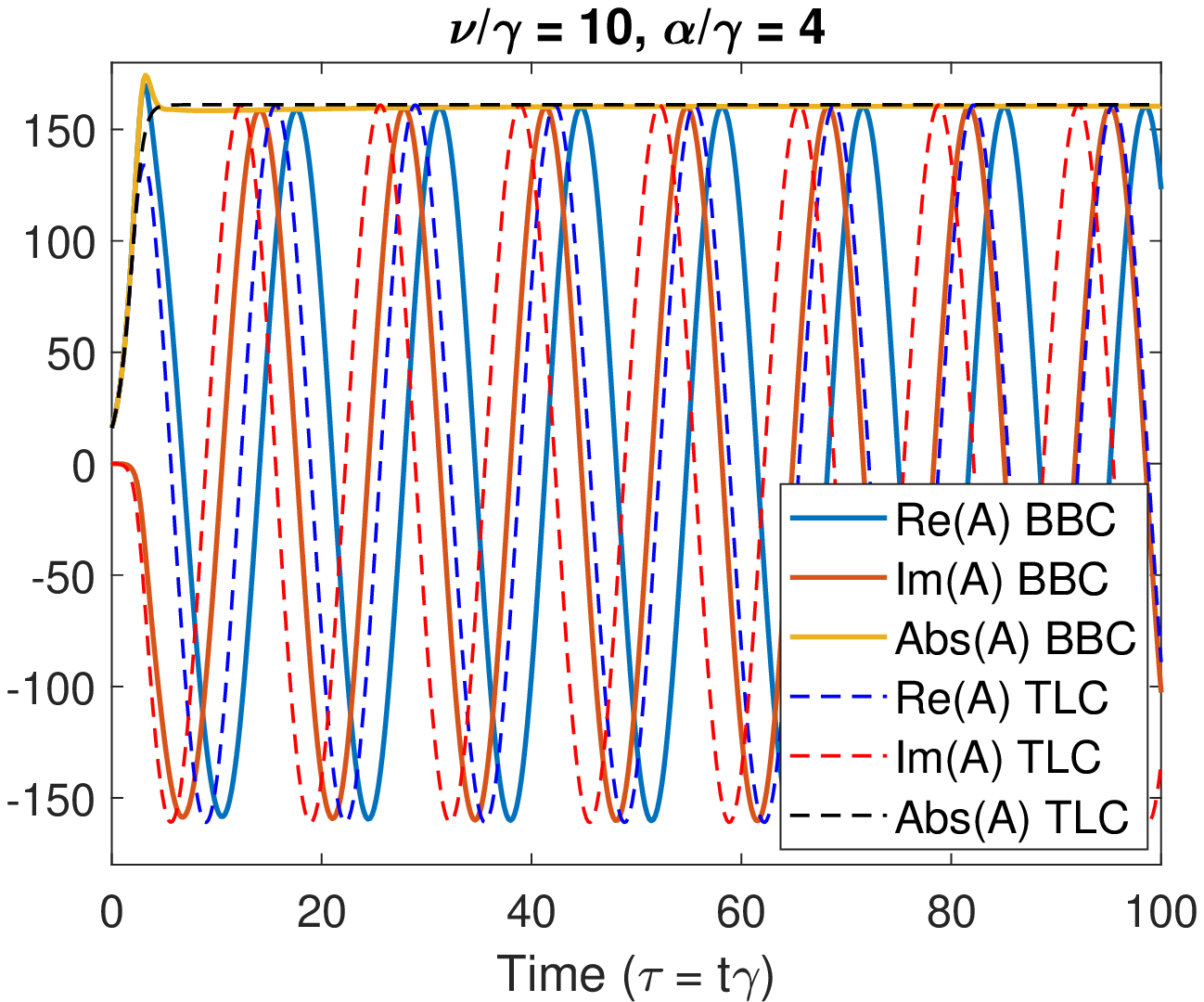}}
\caption{Comparison of the BBC solutions and analytic TLC solutions for different values of $(\alphahat,\nuhat)$. Solid curves result from numerical integration of the BBC equation (\eqref{eq:cubic}) -- blue is $\Re{A}$, orange is $\Im{A}$, and gold is $\abs{A}$. Dashed curves are the analytic TLC solution (\eqref{eq:analytic}, valid when $\nuhat \gg 1$). Top row: fixed ratio $\rrat = 1/3$ with varying (a) $\nuhat = 3$, (b) $\nuhat = 10$, and (c) $\nuhat = 30$. Bottom row: fixed value $\nuhat = 10$ with varying (d) $\alphahat = 1$, (e) $\alphahat = 2$, and (f) $\alphahat = 4$.}
\label{fig:vary}
\end{figure*}

\section{Comparison of the analytic time-local solution to the numerically integrated Berk-Breizman cubic equation}
\label{sec:cubic}

\subsection{Quasi-steady solutions when scattering is much larger than the growth rate \texorpdfstring{$(\nu \gg \gamma)$}{(nu >> gamma)}}
\label{sec:cubbignu}

The analytic solution to the TLC equation derived in \secref{sec:exact} in the quasi-steady regime will now be compared against the exact solution of the BBC equation, as determined by numerical integration of \eqref{eq:cubic}. This comparison is demonstrated in \figref{fig:vary}. The top row shows solutions for three values of $\nuhat = 3$, 10, 30 with fixed $\rrat = 1/3$. In each plot, the numerically calculated time evolution of $\Re{A(\tau)}$, $\Im{A(\tau)}$, and $\abs{A(\tau)}$ are shown by the solid blue, orange, and gold curves, respectively. The corresponding real, imaginary, and absolute value of the analytic solution are represented by dashed blue, red, and black curves, respectively. 

Because a purely real initial perturbation was used, the amplitude remains real during the linear phase. Typically, the initial growth of the wave will overshoot the asymptotic saturation amplitude before converging to a steady value. For $\nuhat = 3$ (\figref{fig:nuvary-3}), there is a substantial oscillation of $\abs{A}$ (solid gold curve) in the nonlinear phase before the oscillations are damped away and a steady state magnitude is achieved. The magnitude of the analytic TLC solution (dashed black curve) can not capture these oscillations, as they occur on a fast timescale that was washed out in the time-local approximation. Nonetheless, the oscillations are always centered around the TLC solution. A similar phenomena can be seen in the real and imaginary parts of $A$, shown as the blue and orange solid curves, respectively, for the BBC solution. At early times (for instance, $\tau < 20$), there are irregular oscillations in the real and imaginary parts on top of the more regular, slower oscillations that dominate at later times. The slower oscillations are precisely the drag-induced modulations that were derived in \secref{sec:exact}, shown in the dashed red and blue curves. The saturation magnitude increases substantially in the three cases shown due to the dependence of $\Asat \like \nuhat^2$, as shown in \eqref{eq:asat-approx}. For smaller values of $\nuhat$ than shown, the oscillations in $\abs{A}$ can become larger in magnitude, chaotic, or even lead to divergence in finite time. These solutions are not the primary focus of this paper, but are discussed briefly in \secref{sec:cubsmallnu} for contrast. 

Solutions for $\nuhat = 10$ and $\nuhat = 30$ are shown in \figref{fig:nuvary-10} and \figref{fig:nuvary-30}, resectively, for the same ratio $\rrat = 1/3$. As $\nuhat$ is increased, the overshoot of $\abs{A}$ past the saturation level diminishes, as do the associated irregular oscillations in the early nonlinear phase. Instead, the real and imaginary parts of $A$ begin their regular sinusoidal dependence almost immediately, mimicking the analytic solutions derived under the assumption of $\nuhat \gg 1$. While the analytic solutions have the correct magnitude and frequency of oscillation, there is a noticeable phase lag between the numerical BBC solution and the analytic TLC solution. These differences are due to $\ord{1/\nuhat^2}$ effects which were neglected in the TLC equation. While the next order corrections are difficult to construct explicitly, it can be mathematically explained why the analytic solution underestimates the initial overshoot while overestimating the initial phase. Quantitative convergence studies are performed to support this conclusion. These details are given in \appref{app:lagconv}.

Consider now the effect of varying the relative amount of drag via the ratio $\rrat$ for a fixed value of $\nuhat \gg 1$. The numerically integrated and analytic solutions to the cubic equation for $\nuhat = 10$ with $\alpha = 1$, 2, and 4 are shown in the bottom row of \figref{fig:vary}. Since $\nuhat$ was chosen to be much larger than one in each of these figures, the solutions are very smooth without any of the irregular oscillations exhibited in the case with $\nuhat = 3$. As the ratio $\rrat$ is increased, the frequency of oscillations in $\Re{A}$ and $\Im{A}$ also increases, as predicted by our theoretical calculations. Close inspection of \figref{fig:avary-1}, \figref{fig:avary-2}, and \figref{fig:avary-4} reveals that the saturation level does have a slight increase due to increased $\rrat$ as predicted by theory, but it is a small increase due to the small chosen values of $\rrat$. Overall, very good agreement is found between the analytic TLC solution and numerically integrated BBC solution. 

\subsection{Dynamical solutions when scattering is comparable to the growth rate \texorpdfstring{$(\nu \like \gamma)$}{(nu ~ gamma)}}
\label{sec:cubsmallnu}

While the main focus of this paper is to examine the quasi-steady regime where $\nuhat \gg 1$ since that is where an analytic solution can be derived for the evolution of the mode amplitude, it is still worthwhile to contrast this against cases with more dynamical behavior. When $\nuhat$ is not sufficiently large, \ie when \eqref{eq:nuhatlim} is not well satisfied, additional time scales become relevant to the mode dynamics. Examples of such behavior can be found in \citeref{Berk1996PRL} for the case of the Krook collision operator. The rich nonlinear behavior present at small values of $\nuhat$ was studied in depth and systematically classified in \citeref{Lesur2012NF}. In particular, Fig. 8 of that work shows a transition from steady to non-steady solutions below $\nuhat \approx 2$ for the case of marginal stability, as predicted in \citeref{Lilley2009PRL}. 

\begin{figure}[tb]
\includegraphics[width=\thirdwidth]{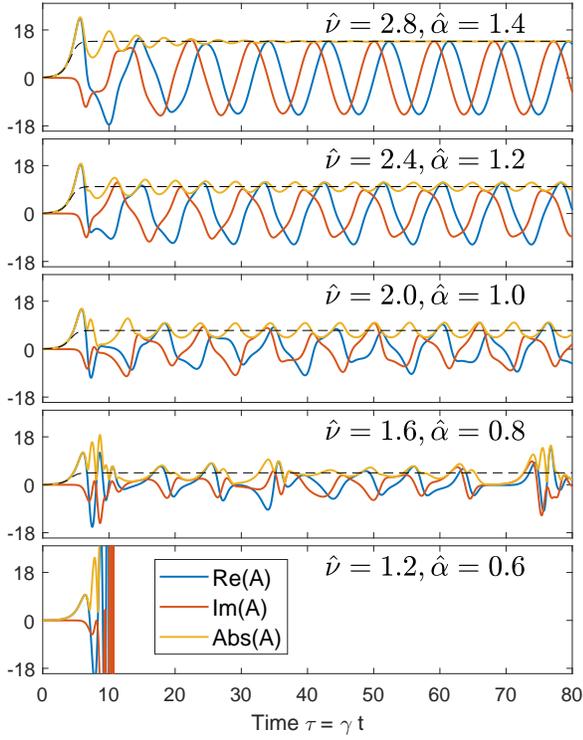}
\caption{Numerical BBC solutions for $\rrat = 0.5$ and varying $\nuhat$ values. In each plot, the blue curve is $\Re{A}$, the orange curve is $\Im{A}$, the gold curve is $\abs{A}$, and the dotted black curve is the analytic TLC solution for $\abs{A}$.}
\label{fig:nu2scan}
\end{figure}

\begin{figure}[tb]
\includegraphics[width=\thirdwidth]{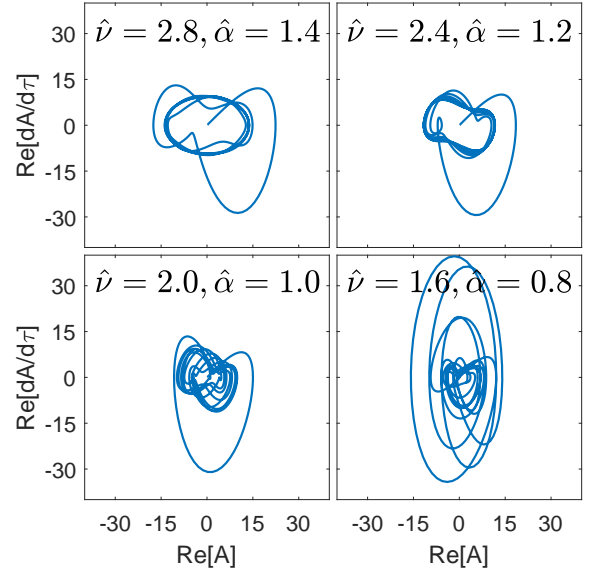}
\caption{Dynamical phase trajectories of $(\Re{A},\Re{dA/d\tau})$ from numerically integrating the BBC equation for $\rrat = 0.5$ and varying $\nuhat$ values.}
\label{fig:nu2ps}
\end{figure}

In \figref{fig:nu2scan}, the BBC equation is evolved numerically for $\nuhat = 2.8, 2.4, 2.0, 1.6, 1.2$ with a fixed ratio of $\rrat = 0.5$. As $\nuhat$ is decreased, the dynamics become progressively more distorted from the quasi-steady solution of the TLC equation. The transition from organized to nearly chaotic behavior can also be seen in the dynamical phase trajectories shown in \figref{fig:nu2ps}, which plots $dA/d\tau$ against $A$. The progressive distortion of the solution to the cubic equation as $\nuhat$ decreases in \figref{fig:nu2scan} is qualitatively similar to that shown in  Fig. 2 of \citeref{Berk1996PRL}. 

At long times, the solution for $\nuhat = 2.8$ is in excellent agreement with the analytic solution to the time-local cubic equation, even though $\nuhat \gg 3$ is not satisfied. The main discrepancy occurs in the early nonlinear phase, as variation on a faster timescale is being damped out, most visible in the magnitude (gold curve). For comparison, the analytic evolution of the magnitude is shown as the black dashed curve. This faster timescale not captured by the time-local approximation also impacts the real and imaginary parts of $A$ (blue and orange curves), which transition from somewhat noisy periodic behavior to regular oscillations around the same time when the magnitude reaches its eventual steady state. In the phase trajectory of $\nuhat = 2.8$ shown in \figref{fig:nu2ps}(a), the long time behavior follows an ellipse in this space, corresponding to periodic motion at a fixed frequency.

As $\nuhat$ is decreased, the evolution becomes more complicated. The case of $\nuhat = 2.4$ is similar to $\nuhat = 2.8$, except that the additional oscillations in the magnitude do not rapidly decay away, resulting in an evolution of the real and imaginary parts that is unmistakably periodic but also somewhat distorted. These distortions are also present in the phase trajectory for $\nuhat = 2.4$, where the clean elliptical trajectory of $\nuhat = 2.8$ is warped in small but noticeable ways. As $\nuhat$ is decreased to 2 and 1.6, the real and imaginary parts of $A$ exhibit progressively more chaotic behavior, which is also reflected in the phase trajectories becoming much more complicated. A Fourier transform shows that additional Fourier components emerge as $\nuhat$ is decreased in this way, leading to multiple spectral lines similar to the explanation given for observations of ``pitchfork-splitting" in terms of the BBC equation with scattering only.\cite{Fasoli1998PRL} Moreover, observed chaotic TAE behavior previously interpreted with the same scattering-only BBC equation\cite{Heeter2000PRL} could also be reproduced by the BBC equation with additional drag, as in the $\nuhat = 1.6$ case. The fact that a collision operator with both scattering and drag could accommodate chaotic solutions was previously demonstrated in 1D Vlasov simulations.\cite{Lesur2012NF} 

Interestingly, the magnitude of $A$ still tracks the quasi-steady solution in each of these cases, indicating that even when the TLC equation is not strictly valid and clearly misses a faster variation of the mode, it still captures the lowest order behavior of the slower timescale for the magnitude evolution. Once $\nuhat$ becomes too small (\eg the $\nuhat = 1.2$ case shown), the BBC equation diverges in finite time even when $\rrat < 0.96$. This is because for $\nu < \nu_\text{crit} \approx 2.05$, steady solutions are possible but not guaranteed to be stable against small perturbations (corresponding to the region between the blue and red curves of the stability diagram in Fig. 1 of \citeref{Lilley2009PRL}). The divergence is not physical, but rather represents a transition from the weakly to strongly nonlinear regime where the near threshold formalism is no longer valid. Dynamical solutions such as frequency sweeping and chirping can occur in this parameter regime.\cite{Duarte2017NF,Duarte2017POP}

\section{Comparison of the analytic time-local solution to nonlinear Vlasov simulations}
\label{sec:bot}

In \secref{sec:cubic}, comparison of the theoretical results against the numerically integrated BBC equation confirmed that the analytic TLC solution was applicable when $\nuhat \gg 1$. However, the BBC equation itself (and therefore its analytic solution in that limit) was derived under the assumption of being close to marginal stability. Hence, it is important to examine how close one must be to marginal stability for these results to be valid. This can be done by simulating the 1D Vlasov system in \eqref{eq:vlasov} directly with the \BOT code,\cite{Lilley2010POP} which takes $\gdgl$ as an input parameter. \BOT is an initial value $\df$ code which represents $\df$ as a finite sum of Fourier harmonics of $kx - \omega t$ in order to simulate the wave-particle interaction due to a single, discrete resonance. Ten harmonics were used for the comparison in this work. The full set of equations solved by \BOT can be found in the appendix of \citeref{Lilley2010POP}.

To compare against the analytic results, \BOT was run for $\nuhat = 10$ with $\gdgl = 0.8, 0.9, 0.95$ while varying the ratio of drag to scattering $\rrat = 0 \tto 0.95$. The time evolution of the real and imaginary parts of $\delta E$ found by \BOT qualitatively reproduce those shown in \figref{fig:vary} for both the analytic TLC solution and numerically integrated BBC equation. 

\begin{figure}[tb]
\subfloat[\label{fig:bot-asat}]{\includegraphics[width = \thirdwidth]{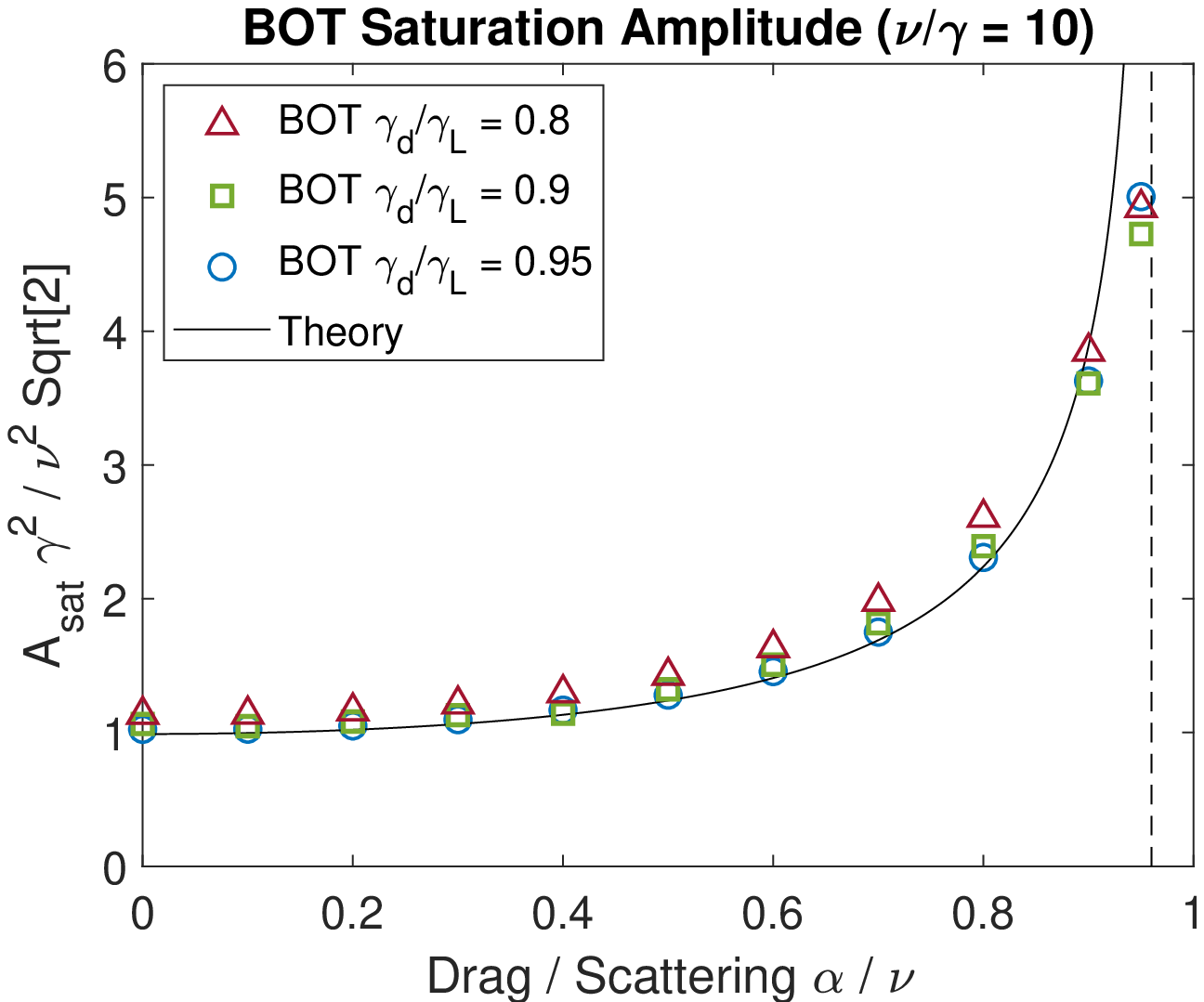}} \\
\subfloat[\label{fig:bot-fsat}]{\includegraphics[width = \thirdwidth]{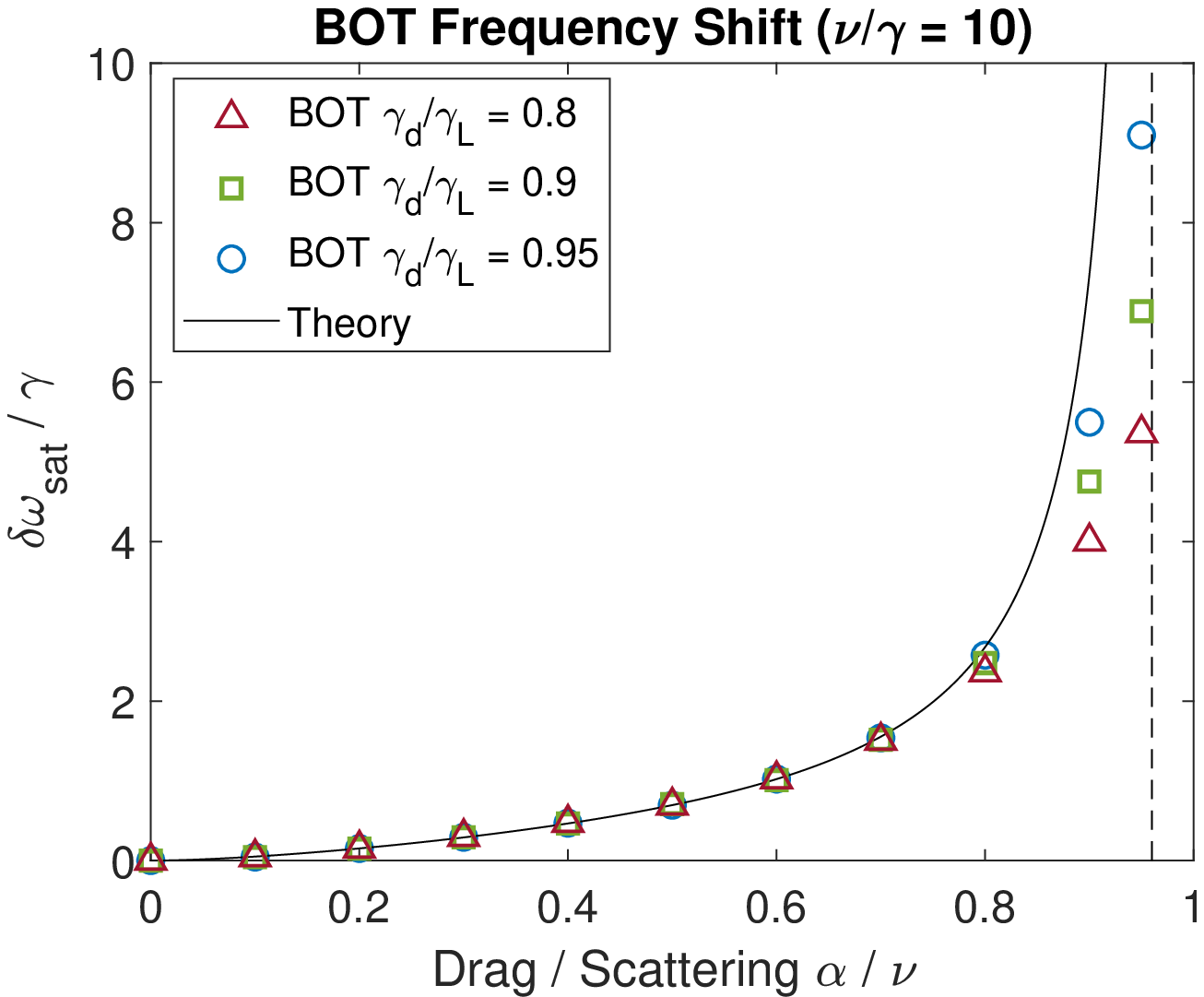}}
\caption{Comparison of \BOT 1D Vlasov simulations to theoretical predictions for (a) saturation amplitude (\eqref{eq:asat}) and (b) frequency shift (\eqref{eq:domegasat}) for different values of marginal stability: $\gdgl = 0.95$ (blue circles), $\gdgl = 0.9$ (green squares), $\gdgl = 0.8$ (red triangles). Black curve gives the theoretical value in the limit of $\nuhat \gg 1$ and $\gdgl \rightarrow 1$. Vertical dotted line indicates the boundary between steady and non-steady solutions. All simulations use $\nuhat = 10$.}
\label{fig:bot}
\end{figure}

The main features to compare quantitatively are the saturation amplitude and frequency shift. These comparisons are shown in \figref{fig:bot}. It is found that neither the amplitude nor the frequency shift is very sensitive to the ratio $\gdgl$ for $\rrat < 0.7$, as there is minimal spread between the results for the simulations with varying $\gdgl$. This suggests that the theoretical results are fairly robust even for cases that are not artificially close to marginal stability. In this range, the agreement between \BOT and theory is extremely close for the frequency shift and moderately close for the saturation amplitude. For the saturation amplitude, the theoretical trend with $\rrat$ agrees with that found in the simulations, but the simulation consistently saturates at a slightly larger value than the theoretical level. As the simulations becomes closer to marginal stability ($\gammad = \gammal$), the saturation level from the simulations becomes closer and closer to the analytic prediction. 

When drag becomes comparable to scattering, the simulation results deviate from the theoretical predictions for both the saturation amplitude and frequency shift. The quantities continue to increase with increasing $\rrat$ in the simulations, but in a less steep fashion than predicted theoretically. This breakdown is due to the BBC expansion parameter $\omegabsq/\nu^2$ becoming larger with $\rrat$ at fixed $\gdgl$. Moreover, the simulations become much more sensitive to the degree of marginal stability in this regime, as the values of both the saturation amplitude and frequency shift spread out significantly for larger values of $\rrat$. The disagreement in the saturation amplitude near $\alpha \approx \nu$ is similar to what was found in a previous comparison between the theoretical prediction and the \code{COBBLES} code (see Fig. A1 of \citeref{Lesur2012NF}). 

These discrepancies are consistent with the insights developed in \secref{sec:validity}. In that section, we demonstrated that the assumption of $\omegabsq/\nu^2 \ll 1$ requires the mode to be closer to marginal stability for larger values of $\rrat$. While $\gdgl = 0.95$ can certainly be considered close to marginal stability, nonetheless it is not sufficiently close for very large values of $\rrat$, as this condition becomes very restrictive near $\alpha \approx \nu$, as shown in \figref{fig:gdglbound}. Although not shown here, the agreement between theory and simulation does improve as $1 - \gdgl$ is decreased further, as predicted. Overall, strong agreement is found between the saturation amplitude and frequency shift derived analytically and measured in the fully nonlinear simulations. 

\section{Discussion of experimental implications for \Alfven Eigenmodes in a tokamak}
\label{sec:experiment} 

While the effects of drag on the evolution of the wave-particle system can be substantial, experimentally isolating this dependence in a tokamak may prove challenging due to the difficulty in 1) accurately calculating the key parameter $\rrat$ and 2) understanding how varying this ratio affects other parameters which the system is sensitive to. These obstacles are explored in this section.

\subsection{Scaling of the ratio of drag to scattering on plasma parameters}
\label{sec:alphaovernuscaling}

The simplest estimate of $\rrat$ can be found from the heuristic arguments\cite{Su1968PRL,Berk1992PRL,Callen2014POP,Catto2020JPP} which give the scaling of the effective rates of scattering and drag within a narrow resonance. 

\begin{align}
	\label{eq:nuheur}
	\nu^3 &\like \nuperp \omega^2 \\ 
	\label{eq:alheur}
	\alpha^2 &\like \tausinv \omega 
\end{align}

Here, $\nuperp$ and $\tausinv$ are the 90 degree pitch angle scattering rate and inverse slowing down time calculated from the following expressions, assuming that both the energetic and background ions are deuterium: 

\begin{align}
\label{eq:nuperp}
\nuperp &= \frac{1}{2}\left(\frac{\vc}{\vres}\right)^3\tausinv \\ 
\label{eq:taus}
\tausinv &= \frac{e^4 m_e^{1/2} n_e \log\Lambda_e}{6(2\pi)^{3/2}\epsilon_0^2 m_p T_e^{3/2}} \\
\label{eq:vc}
\vc^3 &= \frac{3\sqrt{\pi}}{8}\frac{m_e}{m_p}\left(\frac{2 T_e}{m_e}\right)^{3/2} 
\end{align}

These expressions come from \citeref{Goldston1981JCP}, assuming that all ions are deuterium. The constants $m_e$ and $m_p$ are the electron and proton masses, while $\log\Lambda_e \approx 17$ is the Coulomb logarithm. Here $\vc$ is the critical/crossover velocity, above which a beam of fast ions will transfer most of its energy to the thermal electrons instead of ions.\cite{Gaffey1976JPP} To prevent confusion, we will also point out that $\taus$ is the Spitzer slowing down time, not the fast ion slowing down time. These two quantities are related via $\tau_{s,\tep} = (\taus/3) \log(1 + v^3/\vc^3)$.\cite{Stix1972PP,Gaffey1976JPP}

In \eqref{eq:nuperp}, $\vres$ is the velocity of energetic particles at the resonance. In the 1D problem studied in this paper, this is a well-defined quantity. However in true three-dimensional velocity space, resonances depend on all three components of the particle velocity (or equivalently, the three invariants of motion specifying its orbit), so each resonance covers a range of particle velocities in a tokamak. In practice, some resonances may only intersect a slice of the fast ion distribution, in which case assuming a single characteristic value for $\vres$ could be justified. 

Substitution of Eqs. \ref{eq:nuperp} - \ref{eq:vc} into \eqref{eq:nuheur} and \eqref{eq:alheur} yields the following expressions: 


\begin{align}
	\label{eq:nu3}
	\nu^3 \like \left(3427\frac{\text{Hz} \text{ keV}^{3/2}}{10^{20}\text{ m}^{-3}}\right) \frac{\omega^2\neo}{\Wres^{3/2}} \\
	\label{eq:alpha2}
	\alpha^2 \like \left(42.54 \frac{\text{Hz} \text{ keV}^{3/2}}{10^{20}\text{ m}^{-3}}\right)  \frac{\omega\neo}{\Te^{3/2}}
\end{align}

Here we have written $\Wres = m_p \vres^2$ in order to highlight the similar structure of these two expressions. Their combination gives 

\begin{align}
	\frac{\alpha}{\nu} &\approx \frac{1}{\taus^{1/2}\nuperp^{1/3}\omega^{1/6}} \\ 
	&= \left(0.43 \frac{\text{Hz}^{1/6}\text{ keV}^{1/4}}{\left(10^{20}\text{ m}^{-3}\right)^{1/6}}\right)\frac{\Wres^{1/2}\neo^{1/6}}{\Te^{3/4}\omega^{1/6}}
	\label{eq:alphaovernu-happ}
\end{align}

This is where the scaling given in \eqref{eq:alphaovernu-heuristic} originates. More rigorous expressions can be derived directly from the Fokker-Planck collision operator in terms of invariants of motion in a tokamak.\cite{Lilley2009PRL,Lesur2010POP,Duarte2017POP} In reality, the true values of $\alpha$ and $\nu$ require averaging those more detailed expressions (such as given in Eq. 6 and 7 of \citeref{Duarte2017POP}) over both the invariants of motion and orbital trajectories, such as done with the \NOVAK code in previous studies.\cite{Duarte2017NF,Duarte2017POP} For the purpose of illustration, we will consider the simpler heuristic expression contained in \eqref{eq:alphaovernu-happ} in order to understand trends and highlight the opportunities and limitations of applying the theoretical predictions from the 1D bump-on-tail model to an experimental context. 

The first subtlety is that the four parameters appearing in \eqref{eq:alphaovernu-happ} are usually not independent of one another. For instance, the resonance condition establishes a relation between $\Wres$ and $\omega$. For many waves in plasmas $\omega$ will also depend on $\neo$ or $\Te$ through their dispersion relation. To account for this, we will consider the specific example of toroidal \Alfven eigenmodes (TAEs), which are commonly researched due to their potential to generate anomalous fast ion transport.\cite{Heidbrink2008POP,Heidbrink2020POP} 

TAEs have a frequency obeying $\omega \approx \va/2qR$ where $\va = B/\sqrt{\mu_0 m_i n_i}$ is the \Alfven speed and $q = r B_\phi / R B_\theta$ is the tokamak safety factor. They commonly resonate with fast ions at either $\vpar = \va$ or $\vpar = \va/3$. In present-day large aspect ratio tokamaks, fast ions are usually sub-\Alfvenic, so we will assume the mode is driven by the $\vpar = \va/3$ resonance. This implies the relation $\Wres = \mpro\left(\va/3\pitch\right)^2$, where $\pitch = \vpar/v$ is the resonant fast ion pitch. Put together, we find the following expression for $\rrat$ for fast ions resonating with a TAE: 

\begin{align}
	\frac{\alpha}{\nu} \approx 
	\left(0.075 \frac{\left(10^{20}\text{ m}^{-3}\right)^{1/4} \text{ keV}^{3/4}}{\text{T}^{5/6} \text{ m}^{1/6}}\right)
	\frac{B^{5/6}q^{1/6}R^{1/6}}{\neo^{1/4} \Te^{3/4}\pitch}
	\label{eq:alphaovernu-tae}
\end{align}

It's interesting to note that once the TAE frequency dependence is assumed that the scaling with density inverts, demonstrating the importance of varying these parameters in a self-consistent fashion. From \eqref{eq:alphaovernu-tae}, it's clear that the ratio of drag to scattering for TAEs is most sensitive to varying the magnetic field, electron temperature, or the pitch of resonant fast ions. Plugging in typical DIII-D parameters of $B = 2.1$ T, $R = 2$ m, $q = 1.5$, $\neo = 3\ten{19}$ m$^{-3}$, and $\Te = 3$ keV with $\pitch \approx 0.33$ for neutral beam ions gives $\rrat \approx 0.3$. 

\subsection{Dependence of the saturation amplitude on plasma parameters}
\label{sec:expsat}

With the dependence of $\rrat$ on plasma parameters now demonstrated, we must determine which property of the mode evolution presents the best opportunity for measuring the effect of drag. Three candidates come to mind: 1) the mode saturation level, 2) the time it takes the mode to saturate, and 3) the frequency shift of the mode. All three of these quantities increase when $\rrat$ is increased. The mode saturation level (\eqref{eq:asat}) is a good candidate because it is straightforward to measure so long as we understand how it depends on plasma parameters and how to observationally isolate the effect of drag from other effects. The saturation time (\eqref{eq:tinfl}) is more problematic because it is more likely to be blurred by the time dependence of $\fz'(v)$ since instabilities may begin to grow in experiments before the background fast ion distribution has reached a steady state. Moreover, the saturation time has a weaker logarithmic dependence and also depends on the saturated amplitude, so one would be better off simply analyzing the saturation level. Lastly, the frequency shift (\eqref{eq:domegasat}) could be inferred from spectrograms, except that the shift is likely to be smaller than other sources of uncertainty since $\domegasat \like \gamma$ (see \figref{fig:fsat}) and typically $\gamma/\omega \approx 1\%$. Since our theory assumes that the instabilities are near marginal stability, it's unlikely that $\gamma/\omega$ would much greater than this without also having $\gammal\gg\gammad$. Hence the saturation amplitude seems best suited for our purposes. In un-normalized variables, the saturation amplitude can be written as 

\begin{align}
	\label{eq:desat-app}
	\desat &= \frac{2 \mpro}{ek} \nu^2 I(\rrat) \sqrt{1 - \gdgl} \quad\text{ where} \\ 
	I^{-2}(\rrat) &= \nuhat^4\brvar \\ 
	&= \Re{\frac{1}{2}\int_0^\infty \frac{e^{-2 u^3/3 + i \alpha^2 u^2/\nu^2}}{1 - i\alpha^2 / (\nu^2u)}du}
\end{align}

In other words, $I(\rrat)$ is the function plotted in \figref{fig:asat} (modulo a factor of $\sqrt{2}$), which contains the saturation amplitude's explicit dependence on $\rrat$, which also appears in the derived constraint on $1 - \gdgl$ for the validity of the TLC equation (see \eqref{eq:gdgllim}). It is a strictly increasing function that grows slowly for $\rrat \lesssim 0.6$ and then quickly thereafter, formally diverging as $\rrat \rightarrow 0.96$. In the theoretical analysis of the bump-on-tail problem in \secref{sec:analytic}, we assumed that $\alpha, \nu, \gammal, \gammad$ were all independent parameters that could be chosen at will. In reality, each of these quantities are functions of the underlying plasma parameters, so care must be taken to understand how to vary them in a consistent way. For a TAE, \eqref{eq:nu3} can be re-written as 

\begin{align}
	\nu^2 &= \left(5.83\ten{9} \frac{\text{Hz}^2 \text{ m}^{4/3} \text{ T}^{2/3}}{10^{20}\text{ m}^{-3}}\right) \frac{\pitch^2\neo}{q^{4/3}R^{4/3}B^{2/3}}
	\label{eq:nu2TAE}
\end{align}

The dependencies of $\gammad$ and $\gammal$ are more difficult to simply express in terms of the parameters of interest. A crude approximation is to assume that $\gammal/\gammad \propto \beta_\tep / \beta_i$, where $\beta_s = 2\mu_0 P_s / B^2$ is the ratio of kinetic to magnetic pressure for species $s$. This representation is motivated by experimental database studies that have found that the ratio of fast ion to thermal pressure is a key parameter in determining the amount of \Alfvenic activity in a given plasma.\cite{Fredrickson2014NF} Intuitively, larger fast ion $\beta$ means that there may be more free energy available in the non-thermal fast ion distribution to drive the mode. Meanwhile, there are both analytic expressions\cite{Betti1992PFB,Ghantous2012POP,Pinches2015POP} and gyrokinetic simulations\cite{Aslanyan2019NF,Vannini2020POP} indicating that ion Landau damping is often the primary source of TAE damping, which is proportional to the thermal ion $\beta$, which we will assume is similar to the thermal electron $\beta$ for this exercise. Then we can have the rough relation 

\begin{align}
	\frac{\gammad}{\gammal} \like d\frac{\neo\Te}{n_\tep \avg{\W}}
	\label{eq:gdgl-approx}
\end{align}

Here $n_\tep$ is the fast ion density, $\avg{\W}$ is the average fast ion energy, and $d$ is an unspecified proportionality constant. Hence, Eqs.\xspace\ref{eq:alphaovernu-tae}, \ref{eq:nu2TAE}, and \ref{eq:gdgl-approx} can be substituted into \eqref{eq:desat-app} in order to express the scaling of $\desat$ for a TAE as 

\begin{align}
	\desat \propto \frac{\pitch^2\neo}{q^{4/3}R^{4/3}B^{2/3}} 
	I\left(\frac{B^{5/6}q^{1/6}R^{1/6}}{\neo^{1/4} \Te^{3/4}\pitch}\right)
	\sqrt{1 - d\frac{\neo\Te}{n_\tep \avg{\W}}}
	\label{eq:desatTAE}
\end{align}

\subsection{Sensitivity of saturation scaling to uncertainty in the ratio of drag to scattering}

In \secref{sec:alphaovernuscaling}, we argued that the most efficient way to vary $\rrat$ is to vary $B$, $\Te$, or $\pitch$. Since we are exploring how to isolate the effect of drag on $\desat$, the most useful parameter is one which would strongly affect $\rrat$ but only weakly affect the $\nu^2\sqrt{1 - \gdgl}$ dependence of $\desat$. Hence we can rule out $\Te$ since $\sqrt{1 - \gdgl}$ depends on $\Te$, and our assumption that $1 - \gdgl$ is small (modes are close to marginal stability) means that a small variation in $\Te$ will lead to a large variation in the saturation level due to that term, likely obscuring the contribution from drag. The resonant fast ion pitch $\chi$ is also problematic because the $\nu^2$ dependence on $\chi$ is stronger than the $\rrat$ dependence on it. Also $\chi$ is fixed by the physical beam injection geometry, with only a small set of values available for each machine. Thus the most promising parameter is the magnetic field, which has the bonus of being straightforward to control in an experiment, within hardware constraints. 

Unfortunately, \eqref{eq:gdgl-approx} is only a rough approximation, and $\gammal$ may carry some dependence on $B$ since the fast ion drive for TAEs also depends on the number of \emph{resonant} particles with $\vpres \approx \va/3$, which has not been accounted for here. In theory, one could increase the beam power or voltage to try to maintain a constant $\gammal$ while increasing $B$, though the drive also depends on many other neglected parameters. Hence for the purpose of this section -- to illustrate scaling sensitivity to $\rrat$ -- this additional dependence will be ignored. 

\begin{figure}[tb]
\includegraphics[width=\thirdwidth]{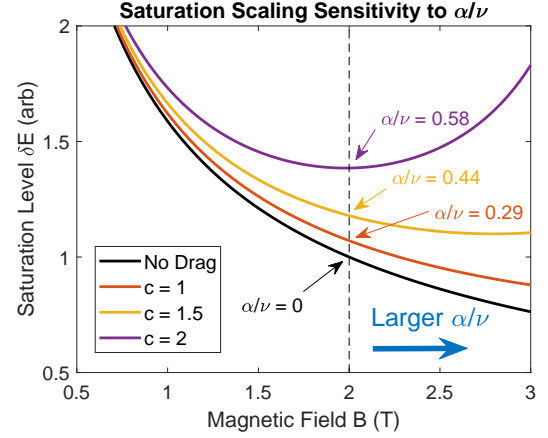}
\caption{Dependence of saturation level on the magnetic field strength with scattering only (black curve) vs scattering and drag when multiplying the heuristic formula for $\rrat$ (\eqref{eq:alphaovernu-tae}) by an ad-hoc coefficient $c$. The labeled values of $\rrat$ are the values for each curve at $B = 2$ T. Increasing $B$ increases $\rrat$ along each curve, except for the black one which has $\alpha = 0$ for the entire curve. The ranges of $\rrat$ for $B = 0.5$ T to 3 T are (0.1,0.4), (0.15,0.6), (0.2,0.8) for the $c = 1, 1.5, 2$ curves, respectively.}
\label{fig:bfield-scaling}
\end{figure}

In the absence of drag, if the magnetic field were increased in a series of otherwise near-identical plasmas (by simultaneously increasing the plasma current to keep $q$ fixed), the mode saturation level would decrease since increasing $B$ decreases $\nu^2$ according to \eqref{eq:nu2TAE}. This dependence is shown as the black curve on \figref{fig:bfield-scaling}, where the values displayed for $\delta E$ are normalized to the value for $B = 2$ T with no drag. On the other hand, increasing $B$ tends to increase $\rrat$, which in turn increases the saturation level due to the increasing function $I(\rrat)$ in \eqref{eq:desatTAE}. So there will be a competition between these two tendencies. For $\rrat \lesssim 0.5$, the $\nu^2$ dependence will determine the overall trend since $I(\rrat)$ has weak dependence for small values of $\rrat$, as shown in \figref{fig:asat}. 

At the nominal magnetic field of $B = 2$ T, the values given at the end of \secref{sec:alphaovernuscaling} yield $\rrat = 0.28$. The saturation level is larger with drag than without drag, but overall the trend is quite similar to the no drag trend. This trend is shown as the orange curve on \figref{fig:bfield-scaling}. Hence in an experiment it would be difficult to isolate the contribution from drag vs the stronger dependence on the $\nu^2 \like B^{-2/3}$ scaling. 

However, as mentioned previously, the evaluation of $\rrat$ is only approximate. We can demonstrate the sensitivity of the $\desat$ scaling to the uncertainty in this calculation by multiplying $\rrat$ by an ad-hoc factor $c$, for instance to account for the missing phase space and orbital averages. Increasing $\rrat$ by a factor of $c = 1.5$, the gold curve shows a larger effect than before, especially at high field. Increasing instead by a factor of $c = 2$, at high field the trend actually reverses. This is because $\rrat$ finally becomes large enough so that the the dependence of $I(\rrat)$ becomes stronger than the $\nu^2$ dependence on $B$. At low field, all of the curves converge to the ``no drag'' scaling because $\rrat \rightarrow 0$ as $B \rightarrow 0$. This demonstrates that while drag can have an important effect, the character of that effect (insignificant, quantitative, or qualitative) is sensitive to the size of $\rrat$, which can not be determined precisely enough with calculations that do not average over phase space and orbital trajectories. 

The main takeaway from this exercise is that although drag can have a nontrivial effect on the mode saturation level, especially when $\alpha$ becomes comparable to $\nu$, in practice it will be difficult to verify this prediction in an experiment since $\rrat$ can not be varied in a truly independent way from $\nu^2$ and $\gdgl$, which also influence the saturation level. Nonetheless, the derived expressions may still be useful in guiding the interpretation of simulations where one has more control over the system's parameters, such as in a recent investigation into the effect of collisions on TAE saturation using the \code{EUTERPE} code.\cite{Slaby2018NF} Moreover, it is crucial to be able to evaluate $\rrat$ accurately in order to determine the overall scaling. The heuristic formula for $\rrat$ is insufficient for this purpose. The example in this section illustrated that varying $B$ to change $\rrat$ from 0.2 to 0.4 will give different behavior than when $\rrat$ changes from 0.4 to 0.8. In reality the phase-space and orbit averages of $\rrat$ are important due to this sensitivity. Properly taking these averages is also essential to determining if a mode will exhibit frequency chirping or steady frequency behavior.\cite{Duarte2017NF,Duarte2017POP,Duarte2018NF} A more promising arena for experimentally verifying the predictions made in this paper might be a basic plasma science facility where it may be operationally more feasible to vary parameters independently of one another. 

\section{Summary and discussion}
\label{sec:conclusions}

\subsection{Summary of main results}
\label{sec:summary}

In this work, the influence of drag on the quasi-steady evolution the 1D bump-on-tail instability was derived analytically and verified against numerical simulations. The previously derived Berk-Breizman cubic (BBC) equation with both drag and scattering collisions\cite{Berk1996PRL,Lilley2009PRL} was used as a starting point, which is a time-delayed, nonlinear integro-differential equation applicable when $\nu^2 \gg \omegabsq$ (which can be satisfied by requiring that the mode be close to marginal stability: $1-\gdgl \ll 1$). Taking advantage of recent analytic insights for the same problem without drag,\cite{Duarte2019NF} the BBC was reduced to a much simpler ordinary differential equation in the additional limit of $\nuhat = \nu/\gamma \gg 1$, referred to as the time-local cubic (TLC) equation. It was rigorously shown that for any value of the ratio of drag to scattering ($\rrat < 0.96$), there exist sufficiently small values of $1 - \gdgl$ and sufficiently large values of $\nuhat$ to justify the reduction of the BBC to the TLC equation. 

An exact analytic solution for the full time evolution of the complex mode amplitude $A(t)$ was subsequently derived for the TLC equation. It was demonstrated that the mode saturation amplitude $\Asat$ is an increasing function of $\rrat$. For $\rrat \ll 1$, the effect on the saturation level is relatively small. But for $\rrat \gtrsim 0.5$ it can become substantial. Moreover, it was found that any non-zero amount of drag leads $A$ to asymptotically oscillate in time, resulting in an apparent frequency shift $\domegasat$ that adds to the linear oscillation frequency of the mode. Although usually small, the frequency shift is also an increasing function of $\rrat$. Interestingly, since the frequency shift only appears once the mode has saturated, this implies that velocity of resonant particles may also shift between the linear and nonlinear phases in order to satisfy the resonance condition throughout the entire evolution of the mode. The integral expressions for $\Asat$ and $\domegasat$ were also evaluated in the regime where $\alpha \ll \nu$ in order to derive more transparent expressions. 

The analytic solution of the TLC equation was compared to the numerically integrated solution of the BBC equation for a range of different values of $\nuhat \gg 1$ and $\rrat$, finding qualitative agreement in all cases and a convergence of the solutions in the limit of $\nuhat \rightarrow \infty$ as predicted. For contrast, numerical solutions of the BBC equation for $\nuhat = 1.2 - 2.8$ were shown to illustrate the transition from irregular to quasi-steady solutions that enables the TLC to capture the mode evolution when $\nuhat \gg 1$. The analytic solutions were also found to be in good agreement with fully nonlinear Vlasov simulations using the \BOT code. Lastly, the prospects for experimental verification of the analytic predictions were considered for toroidal \Alfven eigenmodes in a tokamak. It was determined that although drag can have a non-negligible effect on the observed saturation amplitude, especially when $\rrat \gtrsim 0.5$, observationally isolating the contribution from drag in a tokamak would be challenging since in reality the key parameters $\rrat$, $\nu^2$, and $\gdgl$ can not be treated as truly independent parameters as they are in the simple analytic treatment. 

\subsection{Limitations and disclaimers}
\label{sec:limitations}


Since our analysis took place within such an idealized model to enable analytic progress, there are some key limitations to the conclusions which bear mentioning. First, the Berk-Breizman model which forms the foundation of our analysis only captures the weakest wave-particle nonlinearity (third order in mode amplitude). This is a reasonable assumption for instabilities near marginal stability that will not grow to large amplitudes. However, more strongly driven modes will in reality grow to amplitudes where stronger wave-particle and wave-wave nonlinearities take over. Both theoretical derivations\cite{Hahm1995PRL,Zonca1995PRL,Chen2012PRL,Qiu2016POP} and realistic simulations\cite{Todo2010NF,Todo2012NF} relevant to modern experiments have demonstrated that these higher order nonlinearities can play a role in the saturation of \Alfven eigenmodes in tokamaks. Consequently, we caution that the scenario focused on here -- where collisons, not other mechanisms, control the nonlinear saturation -- is not universal. 





Crucially, as previously emphasized in \secref{sec:experiment}, application of the derived results to a realistic fusion plasma requires one to calculate $\alpha$ and $\nu$ by properly averaging over phase space and resonant orbital trajectories. The heuristic expressions provided in this work for $\rrat$ are useful for cultivating basic intuition but are simply not accurate enough given the fact that the influence of drag hardly matters when $\alpha \ll \nu$ but is very sensitive to the value of $\rrat$ when $\rrat \gtrsim 0.5$. 

In addition, the effective scattering rate $\nu$ which appears in the collision operator should include all forms of scattering. In this work, it was discussed how $\nu \gg \nuperp$ due to the resonances being narrow in velocity space. But scattering from microturbulence\cite{Lang2011POP,Duarte2017NF,Duarte2017POP,Duarte2018NF} and any radio frequency heating\cite{Maslovsky2003PRL,Maslovsky2003POP,Heidbrink2006PPCF,Fredrickson2015NF} present in an experiment can also contribute to $\nu$, potentially affecting the ratio of drag to scattering, and therefore the mode evolution. Incorporating these effects into $\nu$ is beyond the scope of the present work but should be kept in mind when interpreting experiments. 

Furthermore, we analyzed 1D electrostatic waves in a uniform background assuming a constant, infinite background gradient of fast ions. While the same bump-on-tail problem can be formulated in tokamak geometry in terms of action-angle variables,\cite{Berk1997PPR,Berk1997report,Duarte2019POP} the corrections due to background gradients or even cases where the width of the resonance approaches the velocity scale of the variation of the background fast ion distribution would likely lead to some quantitative difference. 

Lastly, we remind the reader that even if a scenario is being considered where none of the above limitations are relevant, care must still be taken to ensure that $1 - \gdgl$ is sufficiently small and $\nu/\gamma$ is sufficiently large such that the TLC equation accurately describes the mode evolution. The constraints on these parameters become more stringent as $\rrat$ increases, with quantitative values given in \figref{fig:bounds}.

\subsection{Discussion and future work}
\label{sec:discussion}

In simple theoretical models, collision operators with only diffusive collisions are often invoked, omitting contributions due to drag based on expectations that $\nu$ may be much greater than $\alpha$. This work shows that the inclusion of drag should be reconsidered, as it can both qualitatively and quantitatively change the nonlinear evolution of marginally unstable waves. Quantitatively, the destabilizing effect of drag which acts to increase the saturation level was previously recognized,\cite{Lilley2009thesis,Lesur2012NF} but most investigations of the effect of drag focused on its role in non-steady behavior (frequency chirping, bursting, chaos, \etc). The apparent frequency shift of quasi-steady solutions due to asymptotic oscillations in the mode amplitude was not previously noted. Moreover, the exact solution of the TLC equation represents a new analytic solution of the bump-on-tail problem (modified by drag) in the limit of large effective scattering ($\nu \gg \gamma$). This provides a potential new verification benchmark for numerical codes. 

The difficulty of experimentally isolating the effect of drag on the nonlinear evolution of the instability should not be construed to mean that the effect of drag can be disregarded. The conclusion is actually the opposite. When drag is small $\rrat < 0.35$, the change in saturated amplitude is relatively small (less than 10\%). But when drag becomes more comparable to scattering, the change in saturation amplitude becomes more substantial, even though the interdependence of $\rrat$, $\nu^2$, and $\gdgl$ unavoidably intertwines their contributions to the saturation level. 

Furthermore, while this paper focused on how drag affects the evolution of the mode amplitude, drag also affects the structure of the resonant-wave particle interaction in phase space and the associated redistribution of fast ions. This aspect of the problem is addressed in \citeref{Duarte2021U}. There it is found that drag shifts the location of the resonance due to breaking a symmetry in the collisional dynamics. Consequently, the perturbed particle distribution $\df$ is sensitive to even small amounts of drag. Since the redistribution of resonant particles is often of direct interest (for instance, for fusion products or other energetic particles transported by \Alfven eigenmodes), one should not conclude that small amounts of drag can be safely omitted from one's fast ion collision operator solely on the basis that the change in mode amplitude is expected to be small based on a heuristic estimate of $\rrat$ from \eqref{eq:alphaovernu-tae}. 

Understanding the influence of drag on the evolution of instabilities offers new avenues for the potential control of \Alfven eigenmodes. While not explored here, one could also consider if any of the available actuators being investigated for stimulating or suppressing \Alfven eigenmodes\cite{Garcia-Munoz2019PPCF} (for instance, targeted ion heating, electron heating, or current drive with externally launched waves) can also be used to effectively modify the ratio of drag to scattering, and therefore the mode saturation amplitude. As an example, experiments on the TJ-II stellarator found a transition from steady state to chirping modes with applied electron cyclotron heating\cite{JimenezGomez2011NF,Nagaoka2013NF} and changes to the rotational transform, \cite{Melnikov2016NF,Melnikov2018NF} demonstrating the possibility of manipulating the critical parameter $\rrat$ to control the effect of drag. 

In the future it would be interesting to address questions which would bridge the gap between the simple model studied in this work and the realistic conditions in fusion plasmas. For instance, comparing the solution of the TLC equation to 3D Vlasov simulations of a modern tokamak. A clear frontier of this line of research is developing a simple evolution equation for waves which are far from marginal stability, which does not currently exist to our knowledge. In a different direction, generalizing the simple analytic model to study neglected but meaningful aspects of the nonlinear behavior, such as zonal flows and resonance overlap, could yield new insights to guide the development of predictive models of realistic \Alfven eigenmode saturation and fast ion transport. 

\section{Acknowledgments}
\label{sec:acknowledgments}

The authors are grateful to M.K. Lilley for making his code \BOT available for public use at the following repository: \url{https://github.com/mklilley/BOT} and to W.W. Heidbrink for helpful feedback. Data used to generate figures available upon request. This research was supported by the U.S. Department of Energy under contracts DE-SC0020337 and DE-AC02-09CH11466.

\appendix 

\section{Derivation of the validity of the time-local approximation}
\label{app:rval}

Consider the BBC equation from \eqref{eq:cubic}, using normalized variables $u = \nuhat z, w = \nuhat x, r = \rrat$ to declutter the analysis. Then it becomes 

\begin{multline}
A'(\tau) = A(\tau) - \frac{1}{2\nuhat^4}\int_0^{\nuhat\tau/2} \int_0^{\nuhat\tau-2u} u^2 e^{-u^2(2u/3+w)}e^{i r^2 u(u + w)} \\
\times A\left(\tau-\frac{u\vphantom{+w}}{\nuhat}\right)A\left(\tau-\frac{u+w}{\nuhat}\right)A^*\left(\tau-\frac{2u+w}{\nuhat}\right) dw du
\label{eq:cubicnorm}
\end{multline}

In deriving the TLC equation (\eqref{eq:tlc}), we approximate the time-delayed mode amplitudes as if they had no time delays. Namely, we take advantage of $A(\tau - \delta/\nuhat) \approx A(\tau)$ when $\nuhat \gg 1$, where $\delta = u, u+w, 2u+w$ each appear in \eqref{eq:cubicnorm}. In this appendix we will derive the condition where this approximation is valid. Specifically, we will require the relative error to be small:  

\begin{align}
	\epsilon &\defined \abs{\frac{A\left(\tau - \frac{\delta}{\nuhat}\right) - A(\tau)}{A\left(\tau - \frac{\delta}{\nuhat}\right)}} \ll 1
\end{align} 

For convenience and without loss of generality, the arguments in the definition of $\epsilon$ will be shifted such that we will actually examine the equivalent expression 

\begin{align}
	\epsilon = \abs{\frac{A\left(\tau + \frac{\delta}{\nuhat}\right) - A(\tau)}{A(\tau)}} \approx \frac{\delta}{\nuhat}\abs{\frac{A'(\tau)}{A(\tau)}} 
	< \frac{3}{\nuhat}\abs{\frac{A'(\tau)}{A(\tau)}} 
\end{align}

\begin{figure}[tb]
\includegraphics[width = \halfwidth]{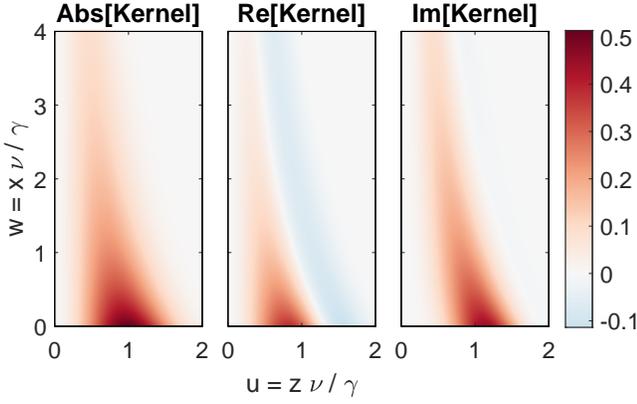}
\caption{Plots of the integral kernel in the BBC equation (\eqref{eq:cubicnorm}) for $\rrat = 0.95$. Left: absolute value. Middle: real part. Right: imaginary part.}
\label{fig:kern}
\end{figure}

The first approximation above is valid for $\delta/\nuhat \ll 1$. Notice that the magnitude of the kernel of the integral, $u^2 e^{-u^2(2u/3+w)}$, peaks at $u = 1, w = 0$, rapidly decaying with characteristic width $\Delta u, \Delta w \approx 1$ away from these values. The absolute value of the kernel, its real part, and its imaginary part, are plotted in \figref{fig:kern} for the extreme case of $\rrat = 0.95$. The magnitude of the kernel does not depend on the ratio $\rrat$. Hence we can bound $\delta/\nuhat \lesssim 3/\nuhat$, since 3 is the maximum value of $\delta \leq 2u + w$ that contributes significantly to the integral. However the decomposition between real and imaginary parts is sensitive to $\rrat$. For small $\rrat \ll 1$, the imaginary part is insignificant and the real part is strictly positive. As $\rrat$ becomes non-negligible, the imaginary part grows while the real part develops a region where it is negative, as is the case shown in \figref{fig:kern}. As will be shown, the real and imaginary parts must be considered separately when evaluating the size of the error in the time-local approximation. 

The next task is to bound the ratio $\abs{A'(\tau)/A(\tau)}$. To do this, decompose $A(\tau) = \abs{A(\tau)}e^{i\phi(\tau)}$. Then 

\begin{align}
	\epsilon < \frac{3}{\nuhat}\abs{\frac{A'(\tau)}{A(\tau)}} = \frac{3}{\nuhat}\sqrt{\left(\frac{\abs{A(\tau)}'}{\abs{A(\tau)}}\right)^2 + \left(\phi'(\tau)\right)^2}
	\label{eq:errgen}
\end{align}

The real and imaginary parts of \eqref{eq:cubicnorm} after this decomposition give 

\begin{widetext}
\begin{multline}
\Aamp' = \Aamp - \frac{1}{2\nuhat^4}\int_0^{\nuhat\tau/2}du \int_0^{\nuhat\tau-2u}dw
u^2 e^{-2u^3/3-u^2 w} 
\abs{A\left(\tau-\frac{u\vphantom{+w}}{\nuhat}\right)}\abs{A\left(\tau-\frac{u+w}{\nuhat}\right)}\abs{A\left(\tau-\frac{2u+w}{\nuhat}\right)} \\
\cos{\left(r^2 (u^2 + uw) 
- \phi(\tau) + \phi\left(\tau-\frac{u\vphantom{+w}}{\nuhat}\right) + \phi\left(\tau - \frac{u+w}{\nuhat}\right) - \phi\left(\tau - \frac{2u+w}{\nuhat}\right) \right)}
\label{eq:cubic-real}
\end{multline}

\begin{multline}
\phi'(\tau) = - \frac{1}{2\nuhat^4 \abs{A(\tau)}}\int_0^{\nuhat\tau/2}du \int_0^{\nuhat\tau-2u}dw
u^2 e^{-2u^3/3-u^2 w} 
\abs{A\left(\tau-\frac{u\vphantom{+w}}{\nuhat}\right)}\abs{A\left(\tau-\frac{u+w}{\nuhat}\right)}\abs{A\left(\tau-\frac{2u+w}{\nuhat}\right)} \\
\sin{\left(r^2 (u^2 + uw) 
- \phi(\tau) + \phi\left(\tau-\frac{u\vphantom{+w}}{\nuhat}\right) + \phi\left(\tau - \frac{u+w}{\nuhat}\right) - \phi\left(\tau - \frac{2u+w}{\nuhat}\right) \right)}
\label{eq:cubic-imag}
\end{multline}
\end{widetext}

Now expand the phase functions $\phi(\tau - \delta/\nuhat) \approx \phi(\tau) - \delta\phi'(\tau)/\nuhat$, which is valid under the same conditions as we used above to approximate $A(\tau - \delta/\nuhat) \approx A(\tau) - \delta A'(\tau) / \nuhat$, namely $\delta/\nuhat \ll 1$. Then to first order in $\delta/\nuhat$, all of the phases in the integrands cancel exactly, resulting in substantial simplification of \eqref{eq:cubic-real} and \eqref{eq:cubic-imag}:

\begin{multline}
\Aamp' \approx \Aamp - \frac{1}{2\nuhat^4}\int_0^{\nuhat\tau/2}du \int_0^{\nuhat\tau-2u}dw
u^2 e^{-2u^3/3-u^2 w} \\
\abs{A\left(\tau-\frac{u\vphantom{+w}}{\nuhat}\right)}\abs{A\left(\tau-\frac{u+w}{\nuhat}\right)}\abs{A\left(\tau-\frac{2u+w}{\nuhat}\right)} \\
\cos{\left(r^2 (u^2 + uw)\right)}
\label{eq:cubic-real2}
\end{multline}

\begin{multline}
\phi'(\tau) \approx - \frac{1}{2\nuhat^4 \abs{A(\tau)}}\int_0^{\nuhat\tau/2}du \int_0^{\nuhat\tau-2u}dw
u^2 e^{-2u^3/3-u^2 w} \\
\abs{A\left(\tau-\frac{u\vphantom{+w}}{\nuhat}\right)}\abs{A\left(\tau-\frac{u+w}{\nuhat}\right)}\abs{A\left(\tau-\frac{2u+w}{\nuhat}\right)} \\
\sin{\left(r^2 (u^2 + uw)\right)}
\label{eq:cubic-imag2}
\end{multline}

In \eqref{eq:cubic-real2}, all terms in the integrand are non-negative except for the cosine term. For the stable steady solutions that we are considering, $r < 0.96$, limiting the size of the argument of the cosine. Although cosine can change sign, inspection of the integrand shows that most of the integrand remains positive when $r < 0.96$, such that the nonlinear term is overall negative. This acts to reduce the growth rate of the mode, leading to the upper bound 

\begin{align}
	\abs{\frac{\abs{A(\tau)}'}{\abs{A(\tau)}}} < 1
	\label{eq:aprimebound}
\end{align}

In other words, the mode's relative growth rate is largest in the linear phase, and decreases as the free energy in the particle distribution is depleted. Rigorously, this argument does not extend to any damped oscillations of $\abs{A(\tau)}$ that may occur in the nonlinear phase around the saturation level. For instance, the case of $\rrat = 1/3$ (see \figref{fig:nuvary-3}) has relatively large oscillations in $\abs{A(\tau)}$ which briefly achieve $\abs{A'(\tau)}/A(\tau) = -1.5$, violating \eqref{eq:aprimebound}. However, such large oscillations do not occur for $\nuhat \gg 3$, which we have already established as a necessary (but possibly insufficient) condition for the validity of our approximation. 

Now consider how we can bound $\abs{\phi'(\tau)}$ by examining \eqref{eq:cubic-imag2}. All terms in the integrand are non-negative except for the sine, which for $r < 0.96$ is also positive almost everywhere that the rest of the integrand is not exponentially small. Hence, $\phi'(\tau) < 0$ at all times. While the mode is growing, $\abs{A(\tau)} > \abs{A(\tau - \delta/\nuhat)}$ leads to a bound 

\begin{multline}
	\abs{\phi'(\tau)} < \frac{\abs{A(\tau)}^2}{2\nuhat^4}\int_0^{\nuhat\tau/2} du \int_0^{\nuhat\tau-2u} dw u^2 e^{-2u^3/3-u^2w} \\ \sin(r^2(u^2+uw))
	\label{eq:phiprimeboundtime}
\end{multline}

In the large $\nuhat \gg 1$ limit, this can be subsequently bounded by constants derived in \eqref{eq:bfull} as 

\begin{align}
	\abs{\phi'(\tau)} < \frac{\bi}{\br}
	\label{eq:phiprimebound}
\end{align}

Once again, this bound does not account for solutions with large oscillations about the saturation level, such as the example previously given, which can transiently exceed the value given in \eqref{eq:phiprimebound} soon after its initial overshoot. A more general upper bound could be written by replacing $\abs{A(\tau)}^2$ with $\abs{A(\tau)}_\text{max}$ instead of $\abs{A(\tau)}_\text{sat}$ which would add a small correction of $\ord{1/\nuhat}$ since it is shown in \figref{fig:lagconv} that $\abs{A(\tau)}_\text{max} - \abs{A(\tau)}_\text{sat} \propto 1/\nuhat$. 

Finally, substitution of \eqref{eq:aprimebound} and \eqref{eq:phiprimebound} into \eqref{eq:errgen} yields the following condition for validity of the TLC equation

\begin{align}
		3\sqrt{1 + \frac{\bi^2}{\br^2}} \ll \nuhat 
		\label{eq:nuhatbound-app}
\end{align}

It is worth repeating that while \eqref{eq:nuhatbound-app} guarantees that the TLC equation will reproduce the BBC equation, there is an independent condition necessary to ensure that the BBC equation itself is valid over the system's entire time evolution. This second condition, as derived in \secref{sec:validity} and reproduced below in \eqref{eq:gdglbound-app} for convenience, comes from requiring that the expansion parameter leading to the cubic equation, $\omegabsq/\nu^2 \ll 1$, is still small even at mode saturation: 

\begin{align} 
1 - \gammad/\gammal \ll \frac{1}{I^2(\rrat)}
\label{eq:gdglbound-app} 
\end{align} 

Here, $I^{-2}(\rrat) = \nuhat^4 \brvar$ depends only on the ratio $\rrat$ and not $\nuhat$ specifically. The numerical values of the conditions in \eqref{eq:nuhatbound-app} and \eqref{eq:gdglbound-app} are shown in \figref{fig:bounds} as a function of $\rrat$. 

\section{Additional properties of the nonlinear evolution}
\label{app:quantities}

Given the analytic expression for the evolution of the mode amplitude, it is useful to calculate some characteristic quantities of the mode evolution. One can calculate the instantaneous nonlinear growth rate $\gammanl$ associated with the time evolution of the wave packet in the $\nuhat \gg 1$ limit via the relation $\Aamp = \Aampz \exp\left[\int_0^\tau \frac{\gammanl(\tau') - \gammad}{\gammal - \gammad}d\tau'\right]$. Comparing \eqref{eq:aamp} to Eq. 3 of \citeref{Duarte2019NF}, we see that they are identical with the replacements $A(\tau)\rightarrow\Aamp, b \rightarrow \br$. Hence, the nonlinear growth rate is similarly given by 

\begin{align}
\Gamma(\tau) \defined \frac{\gammanl(\tau) - \gammad}{\gammal - \gammad} = 1 - \frac{\Aamp^2}{\Asat^2}
\label{eq:gammanl}
\end{align}

Intuitively, the nonlinear growth rate vanishes when saturation is achieved at $\Aamp = \Asat$. Since the inclusion of drag (\eg $\alpha \neq 0$) increases $\Asat$, drag has the destabilizing effect of increasing $\gammanl$ at all points in time relative to its value for $\alphahat = 0$. 

It is also useful to know the characteristic nonlinear saturation time of the wave packet, which can be quantified with the inflection time defined by $d^2\abs{A(\tinfl)}/d\tau^2 = 0$. Differentiation shows that this condition occurs when $\Aamp = \Asat/\sqrt{3}$. The corresponding time is  

\begin{align}
\tinfl = \frac{1}{2}\log\left[\frac{1}{2}\left(\frac{\Asat^2}{\Aampz^2}-1\right)\right]
\label{eq:tinfl}
\end{align}

Interestingly, although drag is destabilizing, it also lengthens the timescale for saturation due to the larger saturation amplitude that it induces. 

It is natural to ask if the solutions we have derived for the TLC equation are stable or unstable to small perturbations. Hence consider a small perturbation to the long time solution of the TLC equation: $A(\tau) = \Aexact(\tau) + \delta A(\tau)$ where $\abs{\delta A} \ll \abs{\Aexact}$. $\Aexact(\tau) = e^{-i\bi\tau/\br}/\sqrt{\br}$ is the previously derived solution and $\delta A = \abs{A_1(\tau)} e^{i\phi_1(\tau)}$ is an arbitrary complex perturbation to the magnitude $\abs{A}$ and phase $\phi$. 

Substitution of $A = \Aexact + \delta A$ into \eqref{eq:aamp} gives $\abs{A_1(\tau)}' = -2 \abs{A_1(\tau)}$ to lowest order in $\abs{A_1/\Aexact} \ll 1$, meaning that small perturbations to the magnitude would decay as $e^{-2\tau}$. Using \eqref{eq:phitime} yields $\phi_1'(\tau) = -2\left(\domegasat/\gamma\right)\abs{A_1}/\abs{\Aexact} \propto e^{-2\tau}$ as well. Hence we have proven that the time-local cubic equation is stable against small perturbation, since any perturbation will rapidly decay in magnitude and its phase will stop evolving in time. This is reasonable since we intentionally derived the TLC equation in the parameter regime where the BBC equation had previously been shown to admit stable steady state solutions.\cite{Lilley2009PRL} One noteworthy finding is that perturbations to the solution of the TLC equation decay without any oscillation in the magnitude. In contrast, the eigenvalues resulting from stability analysis of the BBC equation have nonzero imaginary parts, analogous to an underdamped oscillator. 

As it turns out, the TLC equation is actually robust against perturbations of \emph{any} size. To see this, consider the evolution equation for the evolution of the mode amplitude: $\abs{A}' = \abs{A}\left(1 - \abs{A}^2/\Asat^2\right)$. Imposing a perturbation on the solution is equivalent to choosing an initial condition above the saturation level. For $\abs{A} > \Asat$, clearly $\abs{A}' < 0$ and when $\abs{A} < \Asat$, $\abs{A}' > 0$. Hence any perturbation with $\abs{A} > \Asat$ will monotonically decay until $\abs{A} = \Asat$, without undershooting the unperturbed saturation level. 

When $\rrat > 0.96$, the TLC equation features a finite time divergence which can be readily calculated. This is not a physical divergence, but rather corresponds to the non-existence of steady state solutions, indicating that neglected higher order nonlinearities would become important. While the divergent solutions will necessarily violate the assumed ordering $\omegabsq \ll \nu^2$, they can nonetheless capture the correct behavior at early times, up until this relation breaks down. When divergence occurs, the nonlinear term dominates the linear term in the TLC equation, so \eqref{eq:tlc} further reduces to $\Aamp' = -\br \Aamp^3$ with $\br < 0$ for solutions that diverge (non-steady regime). This new ODE has the simple solution $\Aamp = \Aampz / \sqrt{2\br\Aampz\tau + 1}$, yielding growth like $1/\sqrt{\tdiv - \tau}$ near divergence, where $\tdiv$ is the finite time when blowup occurs. Its value is calculated as 

\begin{align}
	\tdiv = \frac{1}{2}\log\left(1 - \frac{1}{\br\Aampz^2}\right)
\end{align} 

Note that this quantity is always positive since $\br < 0$ for $\rrat > 0.96$. This gives the characteristic time for transition from the linear to nonlinear phase in cases where the BBC equation does not admit a quasi-steady solution. In contrast to the $\left(\tdiv - \tau\right)^{-1/2}$ divergence found from the TLC equation, the BBC equation with a Krook collision operator previously found a $\left(\tdiv - \tau\right)^{-5/2}$ divergence for $\nu < \nu_\text{crit}$ (with an unspecified $\tdiv$).\cite{Berk1996PRL}

\section{Next order corrections to the time-local solution}
\label{app:lagconv}

The next order effects in $1/\nuhat \ll 1$ can be determined qualitatively in order to understand the primary error between the BBC and TLC solutions. The arguments in this section will adopt the same normalized variables as defined at the beginning of \appref{app:rval}. For illustration purposes, the following arguments will assume $r \defined \rrat \ll 1$. However, the conclusions can be numerically shown to be more general and apply for any $r < 0.96$. 

First consider the evolution of the magnitude, determined by \eqref{eq:cubic-real2}. For $r \ll 1$, the cosine term can be approximated as unity. For times prior to the peak value of $\abs{A(\tau)}$, it is clear that $\abs{A(\tau)}$ is strictly increasing. Hence, approximating $\abs{A(\tau - u/\nuhat)}$ by $\abs{A(\tau)}$ and likewise for the other time delays (which is the approximation made in order to derive the TLC equation) necessarily overestimates the double integral term. This results in an underestimate of $d\abs{A(\tau)}/d\tau$ during these times, leading the lowest order solution in $1/\nuhat$ to miss the amplitude overshoot seen in the numerical solution of the BBC equation. 

For the phase evolution, the argument is essentially the same. Again, assuming $r \ll 1$ and observing that the rest of the integrand is peaked near $u = 1, w = 0$ allows an expansion of sine which does not oscillate in the domain of integration. From \figref{fig:vary}, it is clear that the phase error is an early time effect, also occurring before saturation is reached, in the regime where $\abs{A(\tau)}$ is still strictly increasing. So once again, the time-local approximation which ignores the phase-lags inside the integrand leads to an overestimation of the integral term in \eqref{eq:cubic-imag2} as well. Consequently, the lowest order solution for $\phi'(\tau)$ is too large in magnitude at early times, causing the analytic TLC solution to have a larger complex phase when it starts to grow than the numerical BBC solution. This accounts for the phase lag of the TLC solution behind the BBC solution see in \figref{fig:vary}.

\begin{figure}[tb]
\includegraphics[width=\thirdwidth]{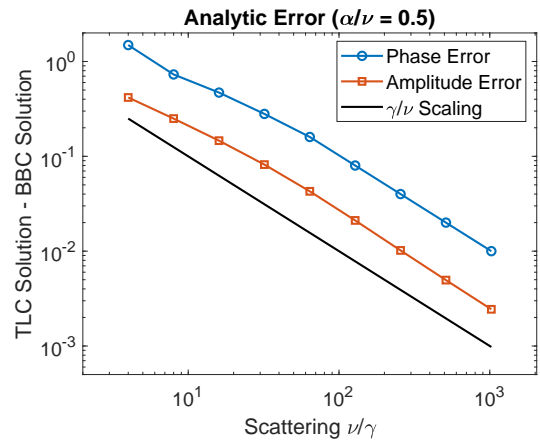}
\caption{Amplitude and phase error between the analytic TLC solution and the numerical BBC solution as a function of the expansion parameter $\nuhat \gg 1$. Blue points indicate the error in the phase. Red points show the error in the peak value of the amplitude. Black line is a reference for $1/\nuhat$ scaling.}
\label{fig:lagconv}
\end{figure}

Moreover, convergence studies were performed to verify that the phase lag and amplitude overshoot are mathematical and not numerical errors. Neither diminishes with smaller time step, so numerical error can be ruled out. Both the error in the phase and the amplitude overshoot between numerical BBC and analytic TLC solutions scale as $1/\nuhat$, as shown in \figref{fig:lagconv}. There, the phase error was calculated as the time difference $(\Delta\tau)$ between the first minimum in $\Im{A}$ for the analytic TLC and numerical BBC solutions using a real initial perturbation. The amplitude error is calculated as the difference between the maximum value of the amplitude achieved by the BBC solution and that of the TLC solution (which also corresponds to the final saturation level). 

\bibliography{all_bib} 

\end{document}